\documentclass[11pt,a4paper]{article}

\usepackage{comment}
\usepackage{float}
\usepackage{color}
\usepackage{multicol}
\usepackage{wrapfig}
\usepackage{graphicx}
\usepackage{bm}
\usepackage{url}
\usepackage{doi}
\usepackage{underscore}
\usepackage{colortbl}
\usepackage{tabularx}
\usepackage{fancyhdr}
\usepackage{ulem}
\usepackage{cite}
\usepackage{amsmath,amssymb,amsfonts}
\usepackage{algorithmic}
\usepackage[]{subcaption}
\usepackage{textcomp}
\usepackage{xcolor}
\usepackage{verbatim}
\usepackage{ascmac}
\usepackage{pdflscape}
\usepackage[margin=25mm]{geometry}

\usepackage{booktabs} 
\usepackage{caption} 
\usepackage{array} 

\hypersetup{
  colorlinks=true,
  linkcolor=blue,
  citecolor=blue,
  urlcolor=blue
}

\title{
  \textbf{Transfer Learning-Based Surrogate Modeling\\
for Nonlinear Time-History Response Analysis\\
of High-Fidelity Structural Models}
}

\author{
  Keiichi ISHIKAWA$^{1}$,
  Yuma MATSUMOTO$^{2}$,
  Taro YAOYAMA$^{1}$,\\
  Sangwon LEE$^{1}$,
  Tatsuya ITOI$^{1,*}$\\
  \text{\small $^{1}$ School of Engineering, the University of Tokyo, Tokyo, Japan}\\
  \text{\small $^{2}$ National Research Institute for Earth Science and Disaster Resilience, Tsukuba, Japan}\\
  \text{\small $^{*}$ Corresponding author Tatsuya ITOI, School of Enginnering, The University of Tokyo, Tokyo, Japan.}\\
  \text{\small itoi@g.ecc.u-tokyo.ac.jp}
}

\date{
  Preprint version: \today
}

\begin{document}

\maketitle

\begin{abstract}
In a performance based earthquake engineering (PBEE) framework, nonlinear time-history response analysis (NLTHA) for numerous ground motions are required to assess the seismic risk of buildings or civil engineering structures.
However, such numerical simulations are computationally expensive, limiting the real-world practical application of the framework.
To address this issue, many previous studies have used machine learning techniques to predict the structural responses to ground motions with low computational costs.
These previous studies typically conduct NLTHAs for a few hundreds ground motions and use the results to train and validate surrogate models.
However, most of the previous studies constructed surrogate models of computationally-inexpensive response analysis models such as single degree of freedom or simple frame models.
Surrogate models of high-fidelity response analysis are required to enrich the quantity and diversity of information used for damage assessment in PBEE.
Notably, the computational cost of creating training and validation datasets increases if the fidelity of response analysis model becomes higher.
Therefore, methods that enable surrogate modeling of high-fidelity response analysis without a large number of training samples are needed.
This study proposes a framework that uses transfer learning to construct the surrogate model of a high-fidelity response analysis model. 
This framework uses a surrogate model of low-fidelity response analysis as the pretrained model and transfers its knowledge to construct surrogate models for high-fidelity response analysis  with substantially reduced computational cost.
As a case study, surrogate models that predict acceleration and inter-story drift ratio at each floor of a 20-story steel moment frame were constructed with only 20 samples as the training dataset.
The responses to the ground motions predicted by constructed surrogate model were consistent with a site-specific time-based hazard.
\end{abstract}

\noindent
\textbf{Keywords : }surrogate modeling, transfer learning, performance based earthquake engineering, time-history response analysis
\section{Introduction}\label{sec:intro}
Future earthquakes are inherently uncertain phenomena and cannot be predicted deterministically in terms of resulting damage and associated frequency.
In rational design decision-making, a framework called performance based earthquake engineering (PBEE)\cite{FEMAp58}\cite{Gunay18082013}\cite{Moehle2004AFM} was proposed to probabilistically evaluate such seismic risk.
The framework consists of the following four steps.
First, the seismic hazard at the site is assessed using probabilistic seismic hazard analysis (PSHA)\cite{Cornell1968} quantified by exceeding probability of ground motion intensity measures (IMs).
Second, engineering demand parameters (EDPs) of structures, such as peak inter-story drift\cite{BAI201196}, peak floor acceleration, and permanent residual displacement\cite{Bojórquez2013} are evaluated by conducting response analyses or by using empirical formulas\cite{Soleimani2022}\cite{Zhong2023}\cite{Chara2011}.
 Third, component damage states are evaluated based on fragility curves, which are functions of the EDPs.
Finally, the consequences, such as fatalities, economic loss and downtime, and their associated likelihoods of occurrence are assessed by integrating the information obtained in the previous three steps, resulting in decision variables.

Although PBEE has been extensively studied and made significant contributions to seismic risk assessment, there is still room for improvement.
In particular, the common use of scalar IMs or EDPs limits the framework, as these quantities fail to capture the essential temporal characteristics of ground motions and responses.
In the first step of current PBEE, or PSHA, a scalar value such as spectral acceleration at the fundamental period \(S_a(T_{1})\) is used.
Then, the evaluations of EDPs are conducted with \(S_a(T_{1})\) as input.
This simplification results in the loss of physically important information inherent in ground motions.
To generate input ground motion time-histories for dynamic response analysis using only limited information such as \(S_{a}(T_{1})\), several ground motion selection methods have been proposed\cite{Jack2011}.
However, ground motions selected using such methods may still exhibit bias in certain characteristics, such as duration.
The advancement of PSHA methodologies has the potential to address these issues by enabling the direct synthesis of ground-motion time-histories directly\cite{Esfahani2022}\cite{Florez2022}\cite{Yamaguchi2024}\cite{Shi2024}.
Matsumoto et al.\cite{Matsumoto2023}\cite{Matsumoto2024}\cite{matsumoto2025} proposed a deep-learning-based ground motion generative model (GMGM) capable of predicting the probability distribution of acceleration time-histories of ground motions, which can be used for Monte Carlo-based time-history analyses of structures.
Conducting time-history response analyses increases the amount of information available for the evaluation of structural damage.
Furthermore, enhancing the fidelity of response analysis models can be another enhancement of performance evaluation.
Instead of using low-fidelity models such as a lumped mass model, utilizing high-fidelity ones can increase the information available in the damage assessment of PBEE. 
However, conducting numerous nonlinear time-history response analyses (NLTHAs), especially utilizing high-fidelity response analysis models with numerous input ground motions, incurs high computational costs.
Therefore, alternative methods are essential for the efficient calculation or prediction of structural responses with reduced computational cost.

Recent studies have attempted to apply machine learning, particularly deep learning, to predict time-history responses of structures with low computational costs\cite{ZHANG201955}\cite{Ni2022}.
For instance, surrogate models that predict time-history responses of a single degree of freedom (SDOF) model, a 2-story simple steel moment frame, and multi-component bridge structures have been constructed using WaveNet, long short-term memory (LSTM), and convolutional neural network (CNN)\cite{NING2023}.
Using attention mechanism, an explainable deep learning model for  time-history response prediction has also been constructed.\cite{SAIDA2025120953}.
Moreover, physics-guided convolutional neural networks (Phy-CNNs), which embed the physical relationships among acceleration, velocity, and displacement, have achieved accurate prediction with a relatively small training dataset\cite{ZHANG2020}.
Although these studies demonstrate the feasibility of deep learning for structural response prediction, most studies focus on low-fidelity or simple response analysis models, where dataset generation through numerous simulations is computationally inexpensive.
However, for high-fidelity models, the cost of conducting hundreds or thousands of NLTHAs for dataset creation is prohibitive.
Thus, there is a need for methods that can construct surrogate models of high-fidelity response analysis models, for which Monte Carlo analysis, using only a limited number of training samples, is impractical in seismic risk assessment.

In this paper, we propose a framework that uses transfer learning to construct surrogate models of high-fidelity response analysis—which are typically computationally expensive to run repeatedly—with a reduced number of training samples.
Transfer learning provides a potential solution when only limited training data are available.
It leverages knowledge obtained from a related source domain, where abundant data exist, and transfers it to a target domain with scarce data\cite{zuhangTL}.
This allows a reduction in the number of samples in the target domain required to construct machine learning models.
It is used in various fields such as image classification\cite{Wang2011}.
While physics-informed approaches have been proposed to reduce the required number of training samples\cite{Ni2022}\cite{ZHANG2020}, the proposed transfer learning framework offers an alternative or complementary data-driven strategy to existing physics-informed approaches for further reducing the number of training samples.

Specifically, the proposed method employs a surrogate model trained on a low-fidelity response analysis model as a pretrained network and transfers the learned knowledge to construct a surrogate model for a high-fidelity response analysis model.
As a case study, surrogate models of a 20-story steel moment frame were constructed by transferring waveform-level temporal representations learned from a low-fidelity SDOF-based surrogate model.
The surrogate models were trained to predict responses to GMGM-based ground motions consistent with a site-specific hazard, using only 20 numerical simulations of a high-fidelity model for training, thereby demonstrating the efficiency of the proposed framework.

The remainder of this paper is organized as follows. Chapter 2 presents the proposed framework that employs transfer learning to construct surrogate models of high-fidelity response analysis models.
Chapter 3 describes a case study, in which surrogate models of 20-story planar steel moment frame (SMF) are constructed to predict time-history responses to site-specific ground motions. 
Chapter 4 provides the conclusions and discusses future research directions.

\section{Proposed Framework}\label{chap:framework}
This section presents a framework that employs transfer learning to construct surrogate models of high-fidelity response analysis models. In this study, the term high-fidelity is used to denote a multi-story nonlinear structural model with higher computational cost and response complexity than the low-fidelity lumped-mass model, rather than an absolute level of modeling detail.
The proposed framework enables the development of surrogate models for computationally expensive simulations that do not require large training datasets from such simulations.

\subsection{Overview of the Transfer Learning--Based Framework}\label{sec:profw}
\begin{figure}[t]
\centerline{\includegraphics[width=0.65\textwidth]{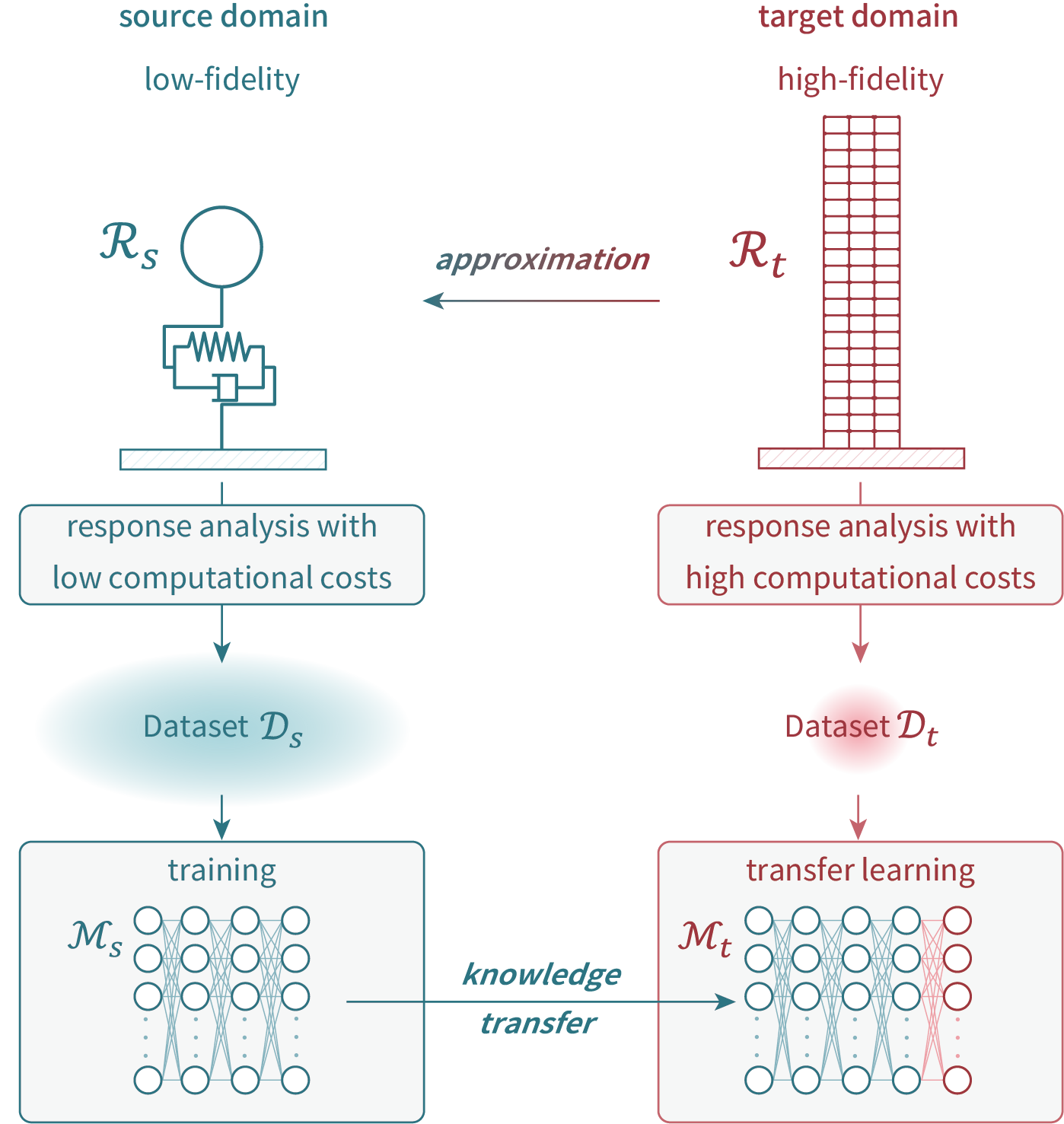}}
\caption{Framework using transfer learning to construct surrogate models of high-fidelity response analysis models. Each dataset contains ground acceleration time-history as input and response analyses results as output.\label{fig:TL_1}}
\end{figure}

The overview of the proposed framework is shown in Figure~\ref{fig:TL_1}, and consists of the following five steps.
\begin{enumerate}
    \item By approximating a target response analysis model (high-fidelity model, \(\mathcal{R}_\mathrm{t}\)), a source response analysis model \(\mathcal{R}_\mathrm{s}\) with lower fidelity is created.
    \item Time-history response analyses of \(\mathcal{R}_\mathrm{s}\) for a large set of input ground motions are conducted and a source dataset \(\mathcal{D}_\mathrm{s}\) is created.
    \item A surrogate model \(\mathcal{M}_\mathrm{s}\) that predicts the seismic responses of \(\mathcal{R}_\mathrm{s}\) is constructed by using dataset \(\mathcal{D}_\mathrm{s}\).
    \item A target dataset \(\mathcal{D}_\mathrm{t}\) is created by conducting a limited number of time-history response analyses of the target response analysis model \(\mathcal{R}_\mathrm{t}\).
    \item Based on the pretrained \(\mathcal{M}_\mathrm{s}\), a surrogate model of the target response analysis model, \(\mathcal{M}_\mathrm{t}\), is constructed by using \(\mathcal{D}_\mathrm{t}\).
\end{enumerate}
In this study, the target surrogate model is constructed in a structure-specific manner. This choice reflects typical PBEE applications, where structural response evaluation is performed on a building-by-building basis.
The source response analysis model \(\mathcal{R}_\mathrm{s}\) created in step 1 approximates \(\mathcal{R}_\mathrm{t}\) by reducing its fidelity while preserving key statistical properties of the seismic responses.
Details of the approximation conducted in this paper are provided in Section \ref{subsucsec:CreSDOF}.

\subsection{Generation of Ground Motions and its Selection}\label{subsec:GMgensel}
\begin{figure}[t]
\centerline{\includegraphics[width=0.6\textwidth]{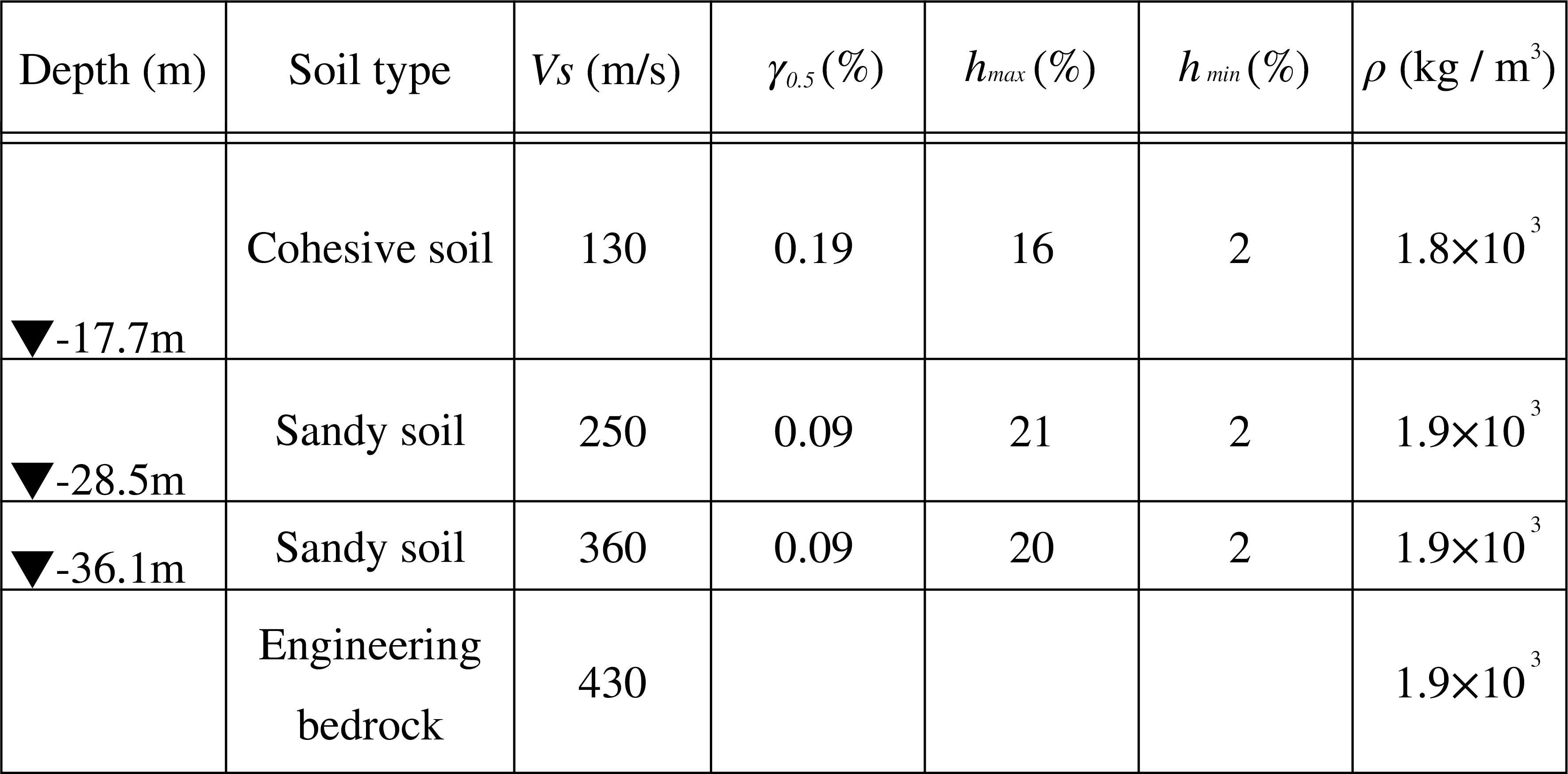}}
\caption{Soil parameters in ground amplification analysis\cite{AIJ2006}.
\(V_{s}\) is seismic shear-wave velocity; \(\gamma_{0.5}\) is standard shear strain; \(h_{max}\) and \(h_{min}\) are the maximum and minimum damping ratios; and \(\rho\) is soil density.} \label{fig:soilparam}
\end{figure}

There are numerous methods for compiling or generating ground motions for NLTHAs.
Notably, the proposed framework is not dependent on the characteristics of the input ground motion.
However, for the case study in this paper, ground motions consistent with site-specific seismic hazard were generated using GMGM\cite{matsumoto2025}.
The GMGM was trained to learn the conditional distributions of observed ground motions and, when embedded within a PSHA framework, enables the generation of synthetic ground motions consistent with seismic hazards. 
GMGM takes a normal random variable as input and outputs a ground motion time-history vector \(\mathbf{g}\in \mathbb{R}^{8192}\) and its label \(\left[M_{w}, R_{RUP}, V_{s30}\right]\), where \(M_{w}\) represents the moment magnitude; \(R_{RUP}\) represents the rupture distance; \(V_{s30}\) represents the average seismic shear-wave velocity in the top \(30~\mathrm{m}\) of the Earth's subsurface.
A stationary Poisson process was used to model the diffuse seismicity of the area with an occurrence frequency of \(0.5\) events per year.
The epicentral area was modeled as a source with a radius of \(55~\mathrm{km}\).
The hypocenter depth was fixed at \(15~\mathrm{km}\).
The lower and upper bounds for the moment magnitude were set to \(M_{w,~\mathrm{min}} = 5.5\) and \(M_{w,~\mathrm{max}} = 6.8\) respectively.
The \(b\) value of Gutenberg-Richter's law was \(b = 0.9\). 
Earthquake occurrence simulations were conductedfor a \(50\)-year period \(10,000\) times using Monte Carlo simulations, generating \(250,476\)  ground motions at the engineering bedrock site, where the \(V_{s30}\) was approximately \(400~\mathrm{m/s}\).
The sampling frequency of $\mathbf{g}$ was 100 Hz, resulting in a duration of 81.92 s, although only the first 40.96 s was used in the case study.
To compute the ground-surface motions, ground motion amplification analyses were performed using the one-dimensional equivalent-linear method \cite{dyneq}, with the soil parameters listed in Figure~\ref{fig:soilparam}.
\begin{figure}[t]
\centerline{\includegraphics[width=0.6\textwidth]{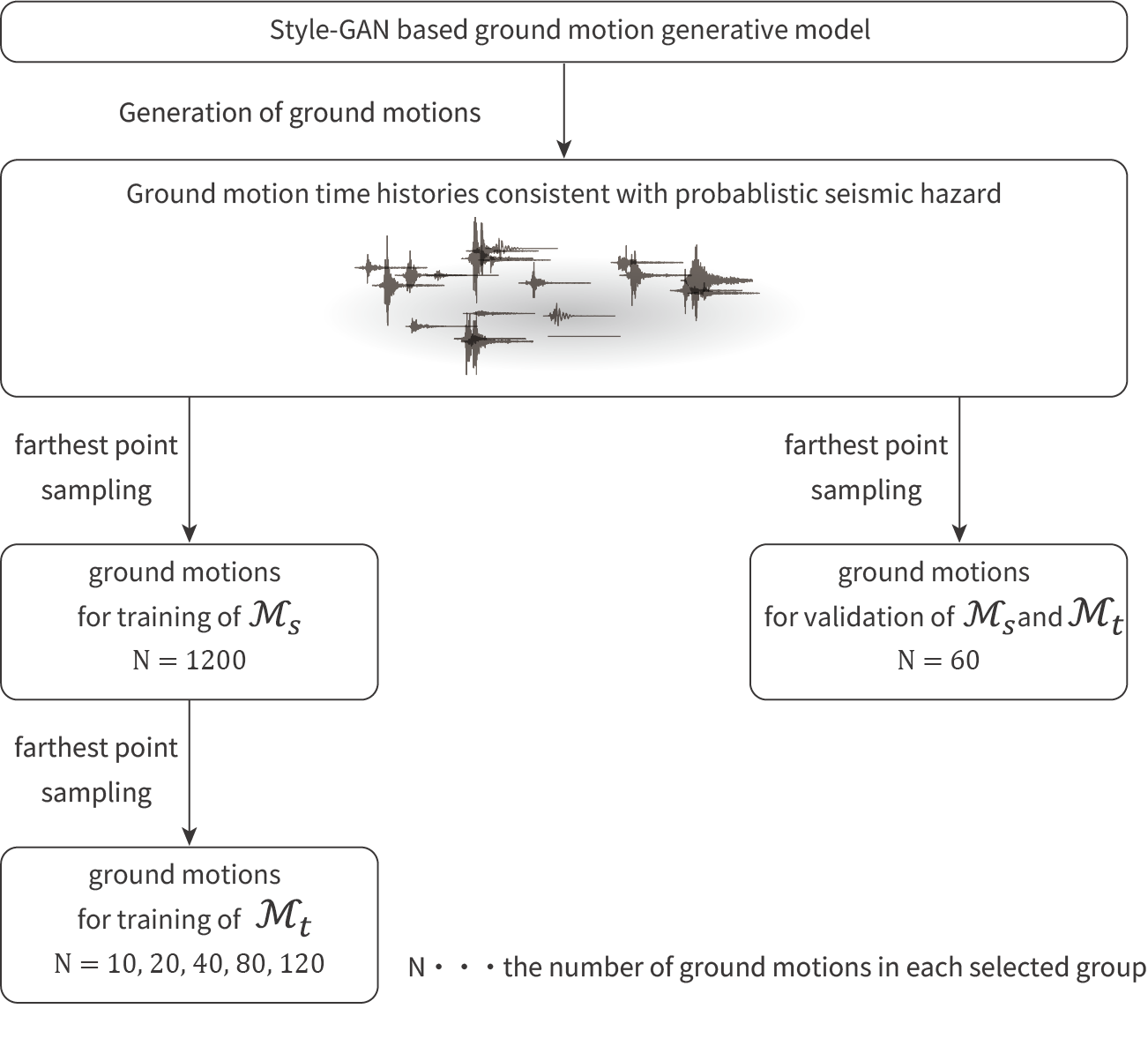}}
\caption{Schematic of the ground motion selection process for training and validation datasets\label{fig:GM_select}}
\end{figure}
\begin{figure}[h!]
\centering
    \includegraphics[width=0.6\columnwidth]{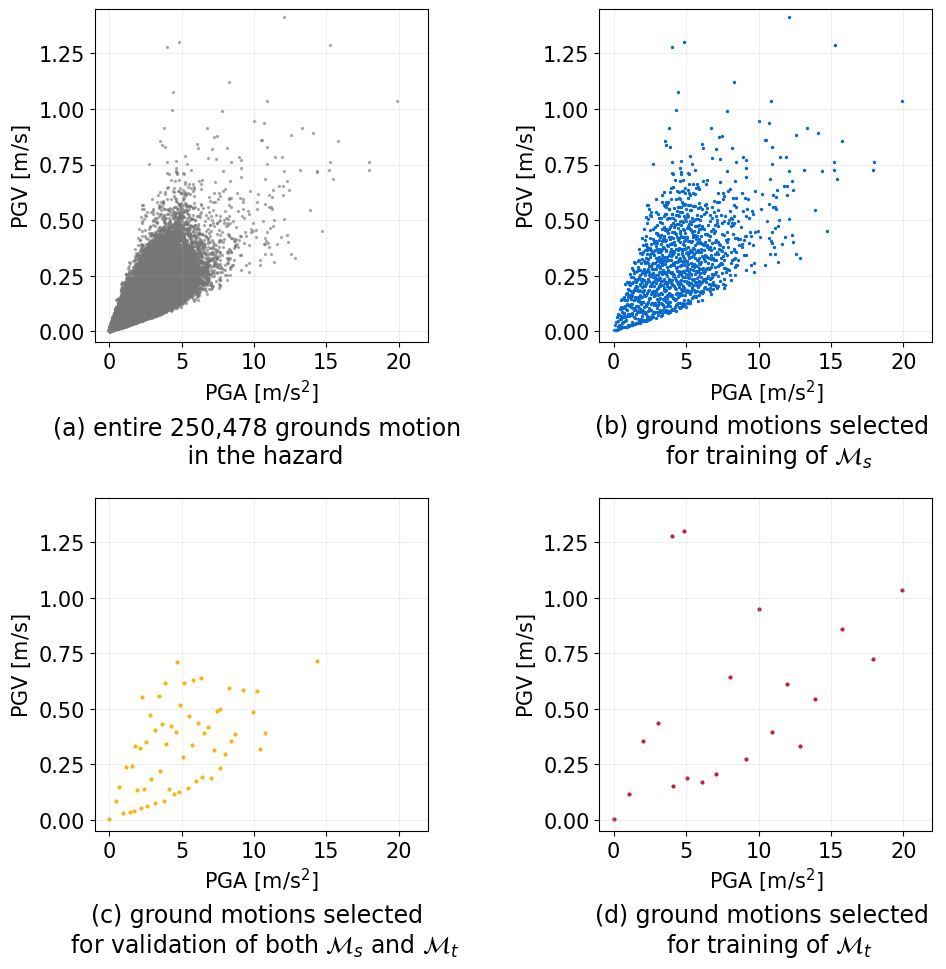}
    \caption{Ground motion distributions of PGA and PGV}
    \label{fig:cs1_GM_select}
\end{figure}

The generated suite contained numerous low-amplitude ground motions originating from frequent, small-magnitude earthquakes and few large-amplitude ones.
Thus, randomly selecting ground motions rarely yielded large-amplitude records in the training dataset, leading to poor predictive performance in these critical large-amplitude regions.
To obtain a set of ground motions that uniformly covers the range from low to high amplitudes, farthest point sampling (FPS) \cite{FS1997} was applied in a two-dimensional intensity-measure space defined by peak ground acceleration (PGA) and peak ground velocity (PGV).
FPS selects ground motions by maximizing pairwise distances in 
PGA-PGV space, thereby producing a dataset that captures diverse amplitude levels.
A schematic of the selection process in the case study is shown in Figure~\ref{fig:GM_select}.
First, \(1200\) ground motions were selected via FPS to create the training dataset for the surrogate model of low-fidelity response analysis model \(\mathcal{M}_\mathrm{s}\). 
Second, another \(60\) ground motions were selected independently via FPS to construct a validation dataset for \(\mathcal{M}_\mathrm{s}\) and \(\mathcal{M}_\mathrm{t}\) evaluation.
Then, \(N\) ground motions were further sampled from the \(1200\) ground motions to compile the training datasets for \(\mathcal{M}_\mathrm{t}\), where different cases of \(N = 10,~20,~ 40,~ 60,~ 80,~120\) were considered  to investigate the effect of the number of high-fidelity simulations.

The PGA and PGV of the selected ground motions are shown in Figure~\ref{fig:cs1_GM_select},indicating that the ground motions cover the low- to large-amplitude regions.
The validation dataset (Figure~\ref{fig:cs1_GM_select}(c)) contained significantly fewer large-amplitude ground motions than the training dataset (Figure~\ref{fig:cs1_GM_select}(b)).
This is because, as illustrated in Figure~\ref{fig:GM_select}, ground motions for the validation were selected only after those for the training had been removed.
Consequently, the validation dataset is insufficient for a comprehensive evaluation of performance under high-amplitude events.
Accordingly, this validation dataset is intended to evaluate baseline waveform-level prediction performance, rather than comprehensive accuracy under high-amplitude ground motions.
To further assess the accuracy of prediction across the entire amplitude range (from low to high), we compared the predicted and true responses using an additional set of \(10,000\) randomly selected ground motions from the the original \(250,476\) ground motions in subsection \ref{sec:comp_surro_res}.

\subsection{Determining the Parameters of \(\mathcal{R}_\mathrm{s}\)}\label{subsucsec:CreSDOF}
In our framework, any response analysis model whose computational cost is lower than that of \(\mathcal{R}_\mathrm{t}\) can be used as the low-fidelity model (\(\mathcal{R}_\mathrm{s}\)).
For example, if the target response analysis model \(\mathcal{R}_\mathrm{t}\) is the steel moment frame model, a simpler model such as an SDOF or multi degree of freedom (MDOF) model can be used as \(\mathcal{R}_\mathrm{s}\).
The case study uses SDOF models with a bilinear shear spring as \(\mathcal{R}_{\mathrm{s}}\) enabling \(\mathcal{M}_\mathrm{s}\) to learn the statistical characteristics of waveforms of the time-history responses.
The characteristics which are not explicitly captured by \(\mathcal{R}_\mathrm{s}\) are intended to be captured in the transfer learning phase.

In transfer learning, it is essential that the source and target datasets share similar characteristics to ensure effective knowledge transfer.
Thus, the time-history data contained in \(\mathcal{D}_\mathrm{s}\) and \(\mathcal{D}_\mathrm{t}\) must have similar characteristics, which means that the time-history responses of \(\mathcal{R}_\mathrm{s}\) and \(\mathcal{R}_\mathrm{t}\) to the same input ground motions should be similar.
To achieve this, the parameters of the SDOF model \(\mathcal{R}_\mathrm{s}\) are determined by minimizing the discrepancy in top-floor relative displacement responses, with the objective of aligning the temporal response characteristics—such as dominant time scales and the onset of nonlinear behavior—between  \(\mathcal{R}_\mathrm{s}\) and \(\mathcal{R}_\mathrm{t}\). This alignment is not intended to enforce equivalence of detailed floor-wise response quantities.

%
The parameters to be optimized are the a natural period \(T_\mathrm{s}\), damping ratio \(\zeta_\mathrm{s}\), yield strength \(f_{\mathrm{y, s}}\), and stiffness reduction ratio \(r_{\mathrm{post}}\), where the mass \(m_\mathrm{s}\) is normalized as \(m_\mathrm{s} = 1~\mathrm{kg}\) to reduce the number of parameters.
These parameters (\(T_{\mathrm{s}}\), ~\(\zeta_\mathrm{s}\), ~\(f_{\mathrm{y, s}}\), ~\(r_{\mathrm{post}}\)) are determined via Bayesian optimization using a black-box optimization framework, Optuna\cite{optuna_2019}.
The objective function is defined as follows.
\begin{align}
  &L = \sum_{i}\frac{1}{T_{\mathrm{step}}}\sum_{t=0}^{T_{\mathrm{step}}}\left(y_{\mathcal{R}_\mathrm{s}, i}(t)-y_{\mathcal{R}_\mathrm{t}, i}(t)\right)^2, ~~\left(t = 0, 1, 2, \cdots, T_{\mathrm{step}}\right)
\end{align}
where \(y_{\mathcal{R}_\mathrm{s}, i}\) is the relative displacement of the SDOF model due to the \(i\)-th ground motion, and \(y_{\mathcal{R}_\mathrm{t}, i}\) is the relative displacement at the top floor of \(\mathcal{R}_\mathrm{t}\).
It should be noted that only a small subset of high-fidelity responses is used to calibrate the low-fidelity model parameters.
For this optimization, \(20\) ground motions were selected from the subset used to create $\mathcal{D}_\mathrm{t}$.


\subsection{Creation of Datasets for Training and Validation}\label{subsubsec:CreDS}
In the case study, \(\mathcal{M}_\mathrm{s}\) is trained to predict time-history responses of relative acceleration, relative velocity, relative displacement and shear force of the lateral spring of the SDOF model (\(\mathcal{R}_\mathrm{s}\)).
Accordingly, the dataset \(\mathcal{D}_\mathrm{s}\) for training and validation is constructed by pairing each input ground-motion time-history with the corresponding response time-histories computed from \(\mathcal{R}_\mathrm{s}\).
Each input ground motion is an acceleration time-history of length \(T_{\mathrm{step}} = 4096\) points (corresponding to \(40.96~\mathrm{s}\) with a \(\Delta t\) of \(0.01~\mathrm{s}\)).
The output response time-history for \(\mathcal{R}_\mathrm{s}\) has a dimension of \(T_{\mathrm{step}}~\times~4\).
For normalization, each input ground motion and the resulting output responses are normalized by dividing by the maximum absolute value observed across the entire \(\mathcal{D}_\mathrm{s}\) dataset.
This normalization ensures that all features have comparable magnitudes.

Similarly, the target dataset \(\mathcal{D}_\mathrm{t}\) contains pairs of input ground motions and the corresponding responses obtained from \(\mathcal{R}_\mathrm{t}\).
Two surrogate models, \(\mathcal{M}_{\mathrm{t}, \mathrm{accel}}\) and \(\mathcal{M}_{\mathrm{t}, \mathrm{IDR}}\), are constructed to predict the relative acceleration and inter-story drift ratio (IDR) on each floor, respectively. Therefore, the output time-history data for \(\mathcal{R}_\mathrm{t}\) has a dimension of \(T_{\mathrm{step}}~\times~20\) for each response type.
Similar to \(\mathcal{D}_\mathrm{s}\), the ground motions and responses in \(\mathcal{D}_{\mathrm{t}, \mathrm{accel}},~ \mathcal{D}_{\mathrm{t},\mathrm{IDR}}\) are normalized.
The input ground motions are divided by the maximum absolute value across the entire \(\mathcal{D}_\mathrm{s}\) and \(\mathcal{D}_\mathrm{t}\) while responses are divided by the maximum absolute value across the entire \(\mathcal{D}_{\mathrm{t},~\mathrm{accel}} \) and entire \(\mathcal{D}_{\mathrm{t},~\mathrm{IDR}}\), respectively.

\subsection{Proposed Architecture of the Surrogate Models \(\mathcal{M}_\mathrm{s}\) and \(\mathcal{M}_\mathrm{t}\)}
\begin{figure}[t]
\centering
    \includegraphics[width=0.6\columnwidth]{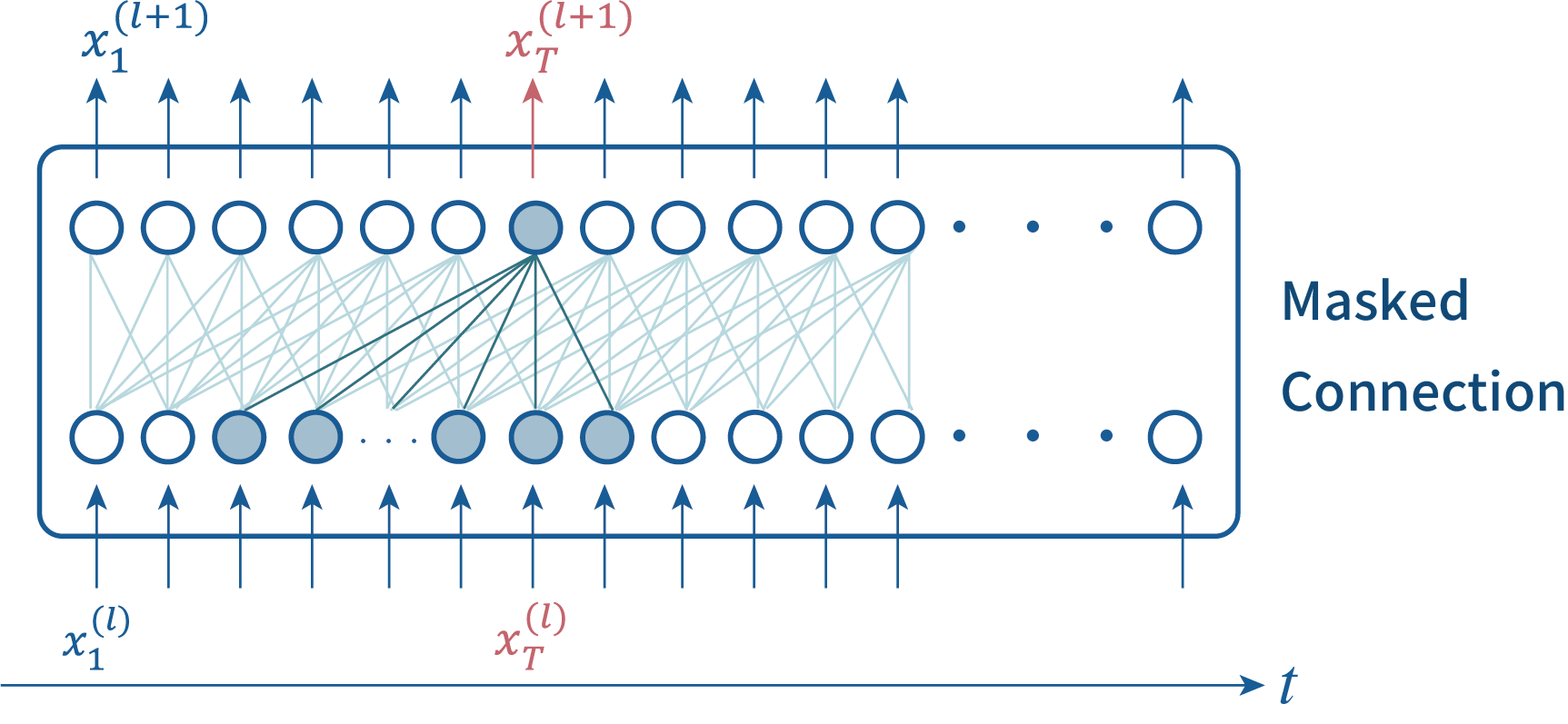}
    \caption{Masked connection used in \(\mathcal{M}_\mathrm{s}\) and \(\mathcal{M}_\mathrm{t}\)}\label{fig:conncet}
    \vspace{5mm}
    \includegraphics[width=0.8\columnwidth]{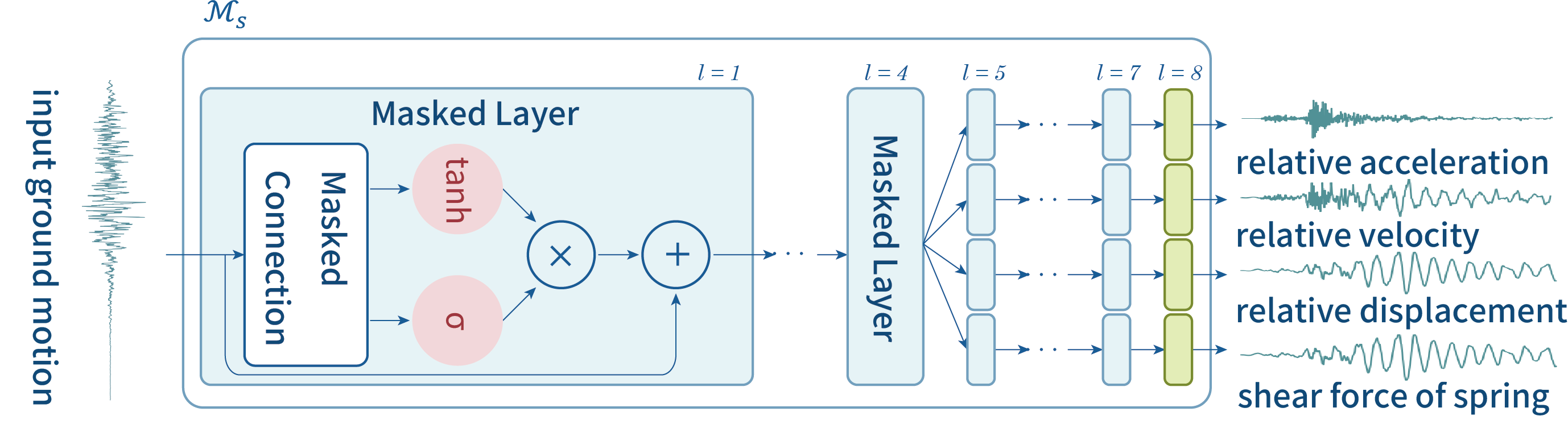}
    \caption{Entire structure of \(\mathcal{M}_\mathrm{s}\)}\label{fig:MS}
    \vspace{5mm}
    \includegraphics[width=0.8\columnwidth]{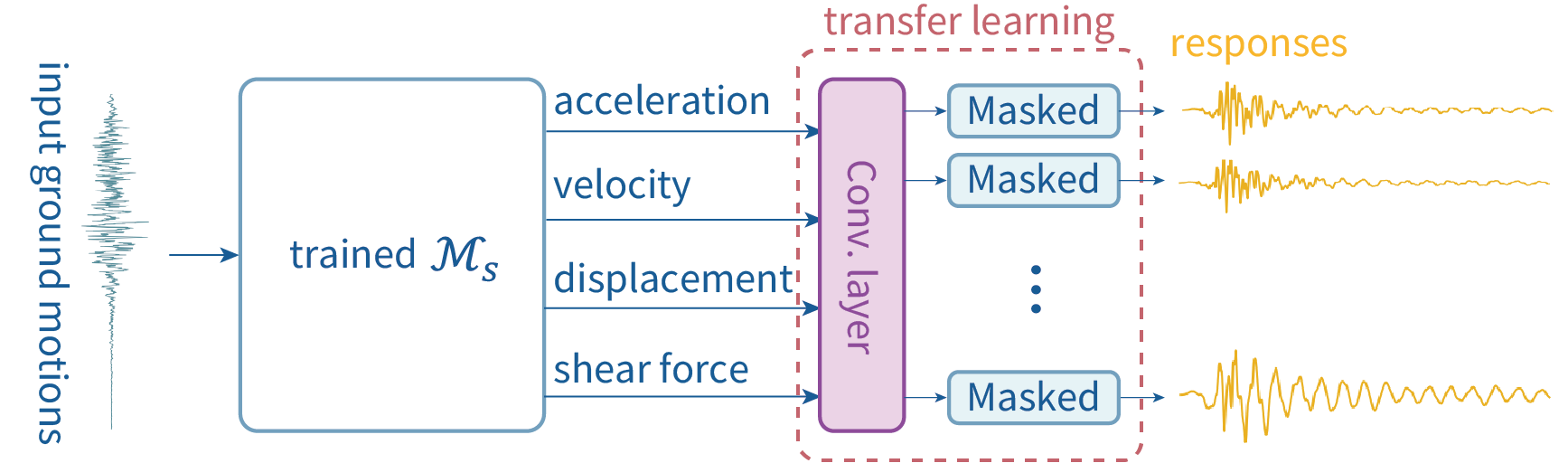}
    \caption{Entire structure of \(\mathcal{M}_\mathrm{t}\)}
    \label{fig:Mtstru}
\end{figure}

For the surrogate models \(\mathcal{M}_\mathrm{s}\) and \(\mathcal{M}_\mathrm{t}\), we introduce a new neural network architecture called masked neural networks (MNN), in which the connections between layers are restricted to reflect the temporal structure of time-history data.
Figure~\ref{fig:conncet} illustrates the architecture of each layer in the MNN.
%
The \(T\)-th output in the \((l + 1)\)-th layer, \(x_T^{l+1}\), depends only on the neighboring outputs in the \(l\)-th layer, that is,
\begin{align}
  x_{T}^{l+1} = \sum_{t=T-T_\mathrm{P}}^{T+T_\mathrm{F}} w_{t,l}~x_{t}^{l}
\end{align}
where \(w_{t,l}\) denotes the \(t\)-th weight in the \(l\)-th layer.
These restricted connections---referred to as masked connections---are implemented by multiplying the weight matrix of each layer by a predefined band matrix.
\(T_\mathrm{F}\) is set to be much smaller than \(T_\mathrm{P}\), to reflect the fact that the response at time $T$ is predominantly dependent on input accelerations in the recent past, thereby improving training efficiency and physical explainability.
Although the masked connections are predominantly past-dependent, a one-step future window (TF = 1) is intentionally introduced. This formulation should therefore be interpreted as a quasi-causal architecture, rather than a strictly causal model.

The entire \(\mathcal{M}_\mathrm{s}\) structure is shown in Figure~\ref{fig:MS}.
The model consists of eight masked layers, with a masked and residual connection to prevent gradient vanishing.
The input to \(\mathcal{M}_\mathrm{s}\) is one-dimensional time-history ground motion data, and the outputs are the four response quantities of interest.
To achieve this, the first four layers have a single channel, whereas the remaining four layers employ four channels.
In the first seven layers (\(1 \leq l \leq 7\)), the past 512 steps and the future single step are used to compute \(x_T^{l+1}\) in the next layer (i.e., \(T_\mathrm{P} = 512, T_\mathrm{F} = 1\)).
For the last layer (\(l = 8\)), highlighted in green in Figure~\ref{fig:MS}, all past steps and the future one step are considered (i.e., \(T_\mathrm{P} = T, T_\mathrm{F} = 1\)) to capture the long-term dependencies that may arise from strongly nonlinear (plastic) responses.

\(\mathcal{M}_\mathrm{t, accel}\) and \(\mathcal{M}_\mathrm{t, IDR}\) are constructed by adding several layers to the trained \(\mathcal{M}_\mathrm{s}\), as shown in Figure~\ref{fig:Mtstru}.
First, a convolutional layer with a kernel size of 2048 and stride of 1 is applied to transform the four-channel outputs from \(\mathcal{M}_\mathrm{s}\) into 20 channels, corresponding to the responses of the 20 stories.
Zero-padding is applied such that the output time-history length remains the same as that of input ground motion, \(T_{\mathrm{step}} = 4096\).
Subsequently, an additional masked layer with \(T_\mathrm{P}=512, ~T_\mathrm{F} = 1\) is added to each channel.
During transfer learning, only the parameters of these newly added layers are updated, while the parameters of the original layers from \(\mathcal{M}_\mathrm{s}\) remain fixed.

\subsection{Training of \(\mathcal{M}_\mathrm{s}\) and \(\mathcal{M}_\mathrm{t}\)}\label{sec:trainMsMt}
Adaptive moment estimation (Adam) to train \(\mathcal{M}_\mathrm{s}\).
Huber loss \(L_{\delta}\) was adopted during the first 50 epochs, adding a physics-informed loss term \(L_{\mathrm{phys}}\) afterwards to improve the accuracy of predicted nonlinear responses, especially relative displacements. It should be noted that the adopted physics-informed loss only serves as a weak physical regularization rather than a rigorous physical constraint.
The total loss \(L_{\mathrm{s}}\) for \(\mathcal{M}_\mathrm{s}\) is formulated as follows:
\begin{align}
  &L_\mathrm{s} = 
  \begin{cases}
  \dfrac{1}{N_{\mathrm{s}}}\sum_{i=0}^{N_{\mathrm{s}}}L_{\delta}(\hat{\mathbf{Z}}_i, ~\mathbf{Z}_i)  & \left(\text{if}~epoch < 50\right) \\
  \dfrac{1}{N_{\mathrm{s}}}\sum_{i=0}^{N_\mathrm{s}}L_{\delta}(\hat{\mathbf{Z}}_i, ~\mathbf{Z}_i) + c\cdot L_{\mathrm{phys}} & \left(\text{if}~50 \leqq epoch \right)
  \end{cases}\\
  &L_\delta(y,\hat{y}) = 
  \begin{cases}
  \dfrac{1}{2}(y-\hat{y})^2, & |y-\hat{y}| \leq \delta \\
  \delta \cdot \big(|y-\hat{y}| - \tfrac{1}{2}\delta\big), & |y-\hat{y}| > \delta
  \end{cases}\\
  &L_{\mathrm{phys}} = \frac{1}{N_s}\sum_{i=0}^{N_s}\sum_{T=0}^{T_{\mathrm{step}}}L_{\delta}\left(\sum_{t=0}^{T}\hat{\dot{y}}_{i}(t)\cdot\Delta t ,~ \hat{y}_{i}(T)\right)
  \label{eq:huber}
\end{align}
Normalized time-history data of each response of the SDOF model calculated by NLTHA \((\mathbf{Z}_{i}\)) and predicted by \(\mathcal{M}_\mathrm{s}\) (\(\hat{\mathbf{Z}}_{i}\)) is expressed as follows:
\begin{align}
  &\mathbf{Z}_i = \left[\frac{\ddot{\mathbf{y}}_i}{\max_t\left|\ddot{\mathbf{y}}_i\right|},
  \frac{\dot{\mathbf{y}}_i}{\max_t\left|\dot{\mathbf{y}}_i\right|},
  \frac{{\mathbf{y}}_i}{\max_t\left|{\mathbf{y}}_i\right|},
  \frac{\mathbf{Q}_i}{\max_t\left|{\mathbf{Q}}_i\right|}\right]\\
  &\hat{\mathbf{Z}}_i = \left[\frac{\hat{\ddot{\mathbf{y}}}_i}{\max_t\left|\ddot{\mathbf{y}}_i\right|},
  \frac{\hat{\dot{\mathbf{y}}}_i}{\max_t\left|\dot{\mathbf{y}}_i\right|},
  \frac{{\hat{\mathbf{y}}}_i}{\max_t\left|{\mathbf{y}}_i\right|},
  \frac{\hat{\mathbf{Q}}_i}{\max_t\left|{\mathbf{Q}}_i\right|}\right]
\end{align}
Here, \(\ddot{\mathbf{y}}_i\), \(\dot{\mathbf{y}}_i\), \(\mathbf{y}_i\) and \(\mathbf{Q}_i\) denote the relative acceleration, velocity, displacement, and restoring force of the SDOF model subjected to the \(i\)-th ground motion, and \(N_\mathrm{s}\) is the number of samples in a mini-batch.
The coefficient \(c\) is introduced as a weighting parameter to regulate the contribution of \(L_{\mathrm{phys}}\) relative to the data-driven Huber loss (\(L_{\delta}\)).
In Eq.~(\ref{eq:huber}), \(\sum_{t=0}^{T}\hat{\dot{y}}_{i}(t)\cdot\Delta t\) represents the numerical integration of \(\hat{\dot{y}}\) with time step \(\Delta t\).
In this study, the coefficient \(c\) is fixed at a constant value of \(5.0\times 10^{-6}\).
In preliminary experiments, this value provided a good balance between data fitting and accurate prediction of nonlinear responses.

Regarding the training of \(\mathcal{M}_\mathrm{t}\), no physics laws were incorporated into the loss function, to examine the effect of transfer learning itself.
\(\mathcal{M}_\mathrm{t}\) was trained using Huber loss, as follows:
\begin{align}
  L_\mathrm{t} = \frac{1}{N_{\mathrm{t}}~F~T_{\mathrm{step}}}\sum_{i=0}^{N_{\mathrm{t}}}\sum_{k=0}^{F}\sum_{T=0}^{T_{\mathrm{step}}}L_{\delta}\left(\frac{y_{i, k, T}}{A_{i,k}}, \frac{\hat{y}_{i, k, T}}{A_{i, k}}\right)
\end{align}
Where \(N_\mathrm{t}, ~F, ~T_{\mathrm{step}}\) are the number of samples, number of floors, and number of time steps, respectively; \(y_{i, k, T}\) is the response for the \(i\)-th sample at the \(k\)-th floor and time step \(T\); \(\hat{y}_{i, k, T}\) is the response predicted by \(\mathcal{M}_\mathrm{t}\).
; and \(A_{i, k}\) is the peak response defined as
\begin{align}
  A_{i, k} = \max_{T}{|y_{i, k, T}|}
\end{align}

\section{Case study | 20-story SMF Subjected to Diffuse Seismicity}\label{chap:casestudy1}
\subsection{Overview of Case Study}\label{sec:overcs1}
The Application of the proposed framework to construct surrogate models of a 20-story SMF.

An SDOF model was used as the source response analysis model (\(\mathcal{R}_\mathrm{s}\)).
Based on the transfer learning scheme, surrogate models that predict relative acceleration and IDR responses at each floor were constructed, considering GMGM-based ground motions compatible with the site-specific time-based seismic hazard described in Section \ref{subsec:GMgensel}.


\subsection{Response Analysis Models \(\mathcal{R}_\mathrm{t}~\mbox{and}~\mathcal{R}_\mathrm{s}\)}\label{sec:20storySMF}
\begin{figure}[t]
  \centering
    \centering
    \includegraphics[width=0.5\columnwidth]{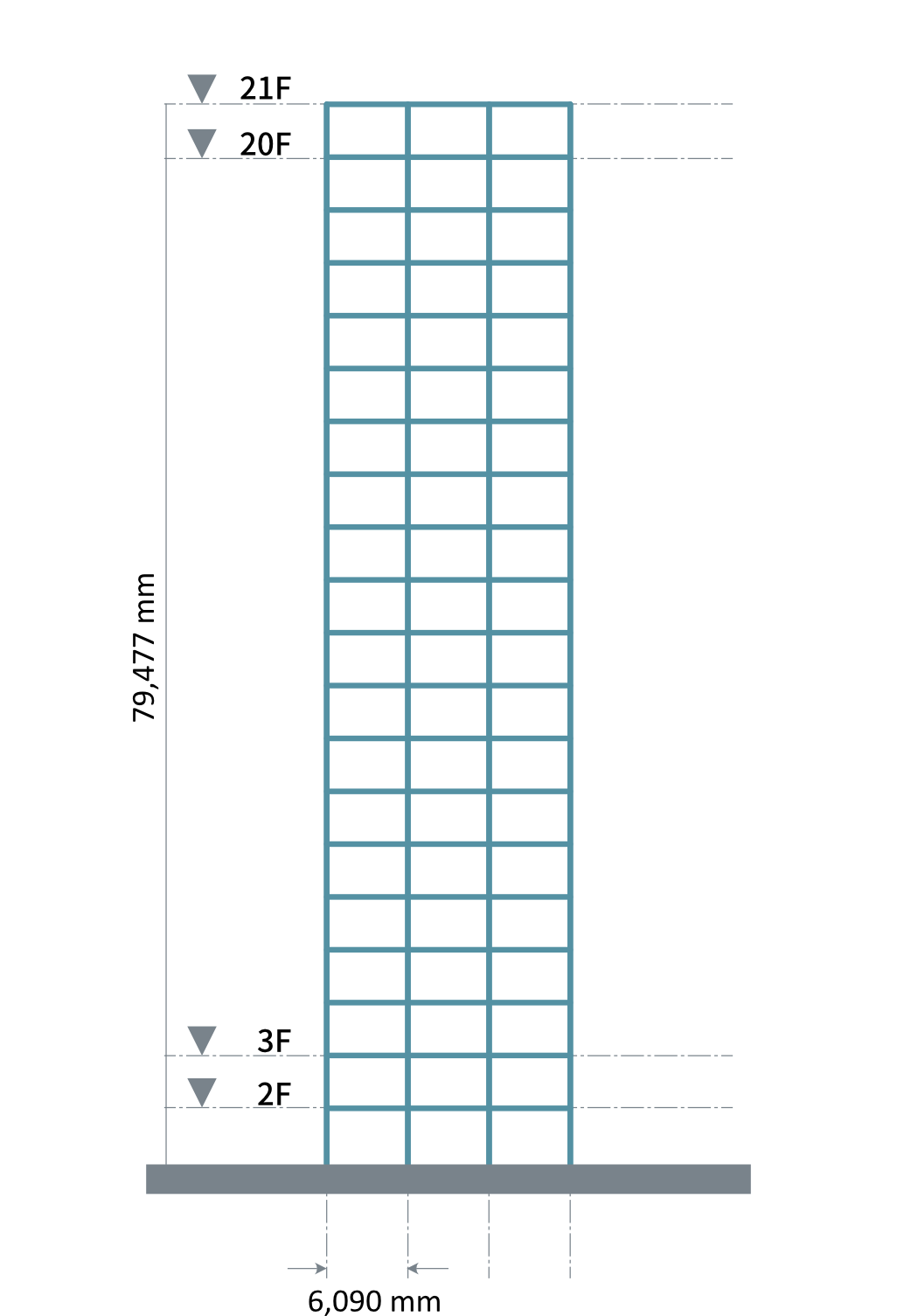}
    \caption{Target response analysis model \(\mathcal{R}_\mathrm{t}\) of the case study, a 20-story planar steel moment frame (SMF)}\label{fig:SMRF}
\end{figure}

The target response analysis model \(\mathcal{R}_\mathrm{t}\) was a 20-story SMF shown in Figure~\ref{fig:SMRF}.
The model was constructed using OpenSeesPy\cite{ZHU20186}.
The sections of beams and columns, spans and heights of each story were defined according to Kircher et al. (2010)\cite{NIST2010} (as detailed in Figure~\ref{fig:SMRF} and Table~\ref{tab:BC_size}).
The SMF in Kircher et al. (2010)\cite{NIST2010} uses reduced beam sections to make the ends of the beams yield earlier than the columns, thereby preventing the column from collapsing and ensuring a favorable strong-column/weak-beam mechanism. 
To incorporate this characteristic into the response analysis model, nonlinear rotational springs were used to connect columns and each ends of the beams, the columns are modeled as elastic.
The modified Ibarra-Medina-Krawinkler (IMK) deterioration model with bilinear hysteretic response\cite{Ibarra2005} was used as the hysteresis model of the beam-end springs.
The parameters of the IMK model were calculated using the formulas listed in the literature\cite{Lignos2011}\cite{peer2010}.
The natural periods of \(\mathcal{R}_\mathrm{t}\) were calculated by eigenvalue analysis, and primary natural period of \(\mathcal{R}_\mathrm{t}\) was \(T_{t, 1} = 2.40~\mathrm{s}\). 
Rayleigh damping was adopted for \(\mathcal{R}_\mathrm{t}\),setting the damping ratios of primary and secondary modes to \(\zeta_{\mathrm{t}, 1} = 0.03\) and \(\zeta_{\mathrm{t}, 2} = 0.03\), respectively.

The parameters of \(\mathcal{R}_\mathrm{s}\) were determined by Bayesian optimization as follows: \(m_\mathrm{s} = 1.0~\mathrm{kg},~T_\mathrm{s} = 2.41~\mathrm{s}, ~f_{\mathrm{y}, \mathrm{s}} =4.33~\mathrm{N}, ~r_{\mathrm{post}} = 0.370,~\mbox{and} ~\zeta_\mathrm{s}= 0.032\).

\begin{figure}
    \centering
    \captionof{table}{Story-wise Beam and Column Sizes\cite{NIST2010}}\label{tab:BC_size}
    \begin{tabular}{ccccc}
      \toprule
      Story & Elevation~(\(\mathrm{m}\))& Beam Size & Exterior Column Size & Interior Column Size\\
      \midrule
      1  & 4.191  & W30X108 & W36X302 & W36X361 \\
      2  & 8.153  & W40X183 & W36X302 & W36X361 \\
      3  & 12.116& W40X215 & W36X247 & W36X361 \\
      4  & 16.078  & W40X215 & W36X247 & W36X361 \\
      5  & 20.041  & W40X215 & W36X231 & W36X361 \\
      6  & 24.003  & W40X215 & W36X231 & W36X361 \\
      7  & 27.965 & W40X277 & W36X231 & W36X361 \\
      8  & 31.928 & W40X277 & W36X231 & W36X361 \\
      9  & 35.890 & W40X277 & W36X231 & W36X361 \\
      10 & 39.853 & W40X277 & W36X231 & W36X361 \\
      11 & 43.815 & W40X297 & W36X231 & W36X361 \\
      12 & 47.777 & W40X297 & W36X231 & W36X361 \\
      13 & 51.740 & W40X297 & W36X231 & W36X361 \\
      14 & 55.702 & W40X297 & W36X231 & W36X361 \\
      15 & 59.665 & W40X264 & W36X182 & W36X302 \\
      16 & 63.627 & W40X264 & W36X182 & W36X302 \\
      17 & 67.589 & W40X211 & W36X160 & W36X231 \\
      18 & 71.552 & W40X211 & W36X160 & W36X231 \\
      19 & 75.514 & W30X108 & W36X160 & W36X182 \\
      20 & 79.477 & W30X108 & W36X160 & W36X182 \\
      \bottomrule
    \end{tabular}
\end{figure}

\subsection{Training and Validation of \(\mathcal{M}_\mathrm{s} ~\mbox{and}~\mathcal{M}_\mathrm{t}\)}\label{sec:DSCRE1}
\begin{figure}[t!]
    \centering
    \includegraphics[width=0.3\columnwidth]{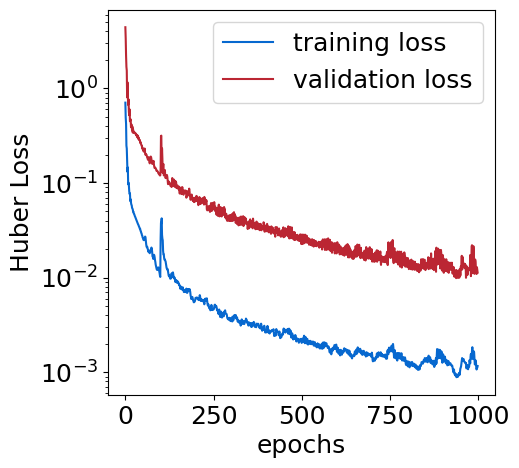}
    \caption{Training and validation loss of \(\mathcal{M}_\mathrm{s}\)}
    \label{fig:cs1_SDOF_loss}
    \vspace{3mm}
\centering
    \includegraphics[width=0.6\columnwidth]{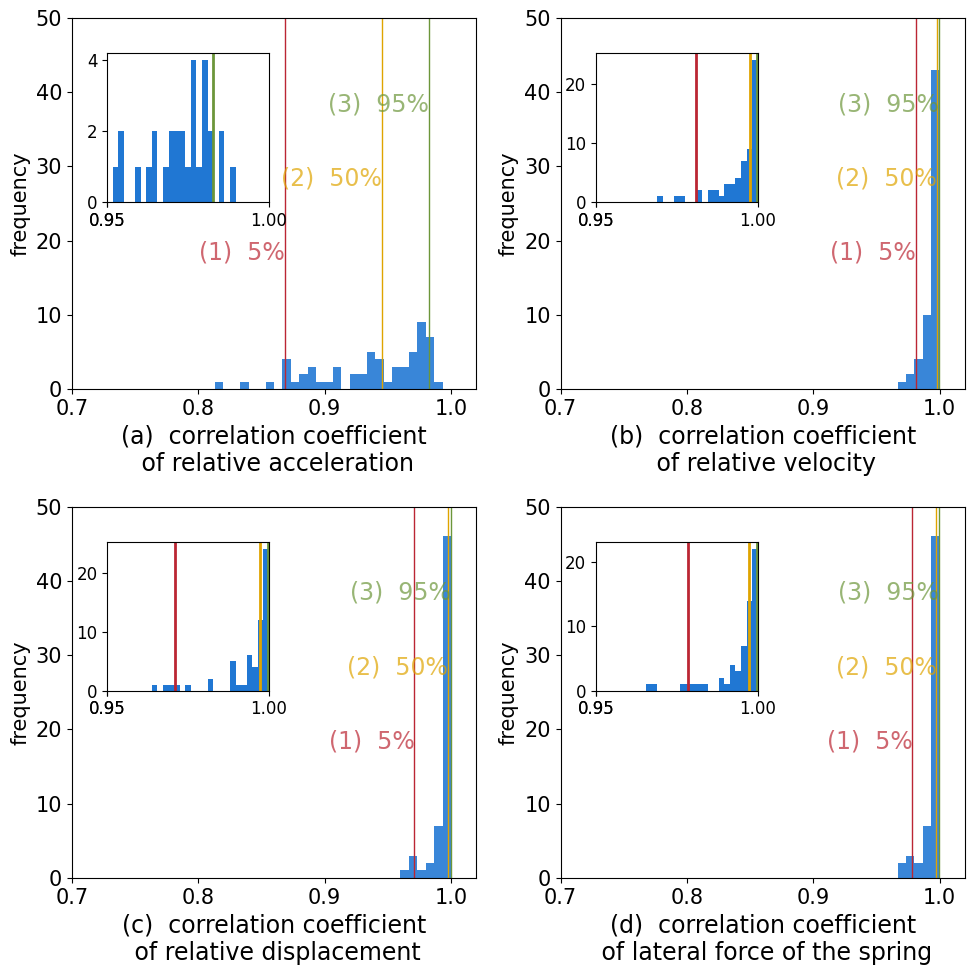}
    \caption{Distribution of \(r_{i, j}\) for validation ground motions}
    \label{fig:cs1_MS_valid_dist}
    \vspace{2mm}
    \includegraphics[width=1.0\columnwidth]{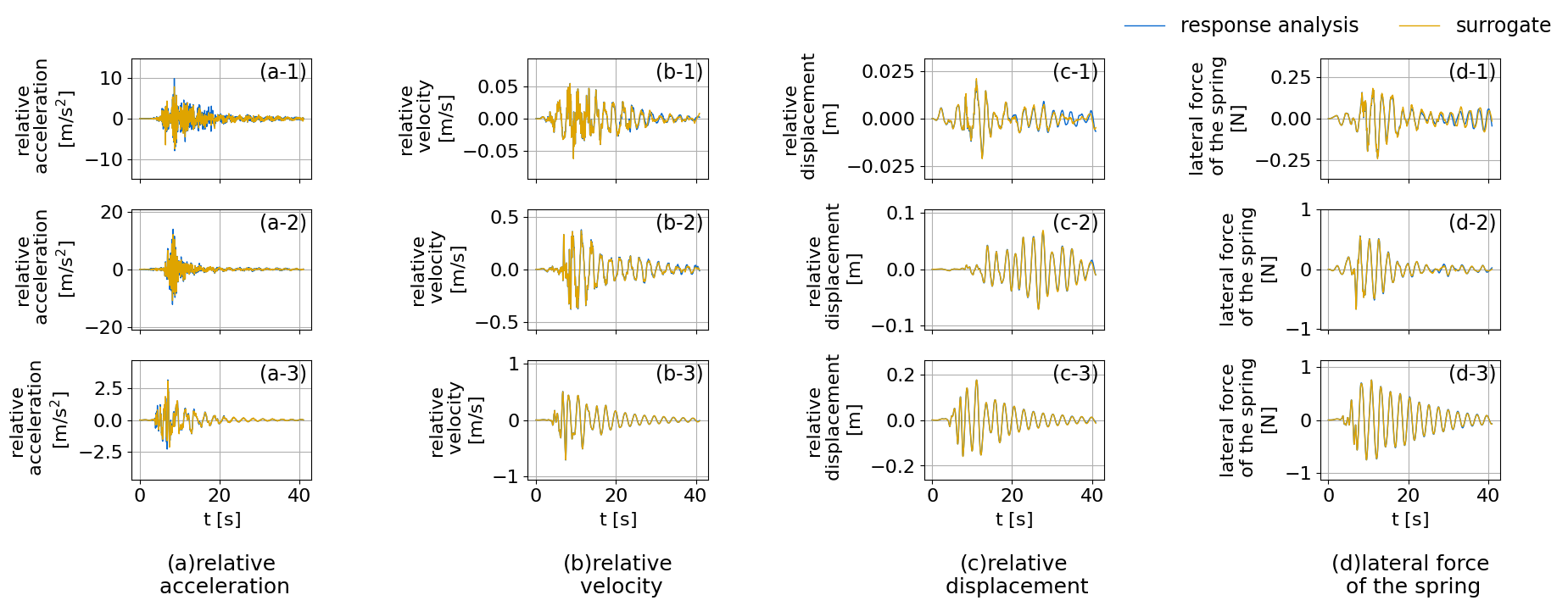}
    \caption{Comparison of response analyses results and prediction by \(\mathcal{M}_\mathrm{s}\) for validation ground motions}
    \label{fig:cs1_MS_valid_wave}
\end{figure}

The NLTHAs of \(\mathcal{R}_\mathrm{s}\) were conducted for the ground motions in the group selected for the training of \(\mathcal{M}_\mathrm{s}\) and validation, thereby creating the training and validation dataset (\(\mathcal{D}_{\mathrm{s} ,\mathrm{train}}, \mathcal{D}_{\mathrm{s}, \mathrm{val}}\)).
Each sample in both \(\mathcal{D}_{\mathrm{s} ,\mathrm{train}} ~\mbox{and}~ \mathcal{D}_{\mathrm{s}, \mathrm{val}}\) has ground motion as input and relative acceleration, relative velocity, relative displacements and shear force of the spring as outputs.
\(\mathcal{D}_{\mathrm{s}, \mathrm{val}}\) contains \(1,200\) samples and \(\mathcal{D}_{\mathrm{s}, \mathrm{val}}\) contains \(60\) samples.
The training of \(\mathcal{M}_\mathrm{s}\) was conducted for \(1,000\) epochs using the Adam optimizer, with the learning rate set to \(5.0\times10^{-5}\).
The result of training is shown in Figure~\ref{fig:cs1_SDOF_loss}.
To validate the prediction performance of \(\mathcal{M}_\mathrm{s}\), the correlation coefficient \(r_{i, j}\) between the ground-truth time-history response \(\mathbf{y}_{i, j}\) and predicted response \(\hat{\mathbf{y}}_{i, j}\) was computed for each ground motion in \(\mathcal{D}_{\mathrm{s}, \mathrm{val}}\) as

\begin{figure}[t!]
\centering
\begin{minipage}[b]{0.45\linewidth}
    \centering
    \includegraphics[keepaspectratio, scale=0.4]{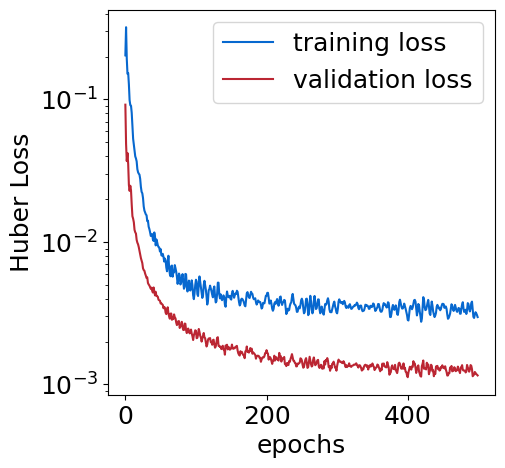}
    \subcaption{Loss function of \(\mathcal{M}_{\mathrm{t},~\mathrm{accel}}\) trained by the dataset that include \(20\) samples}\label{fig:cs1_accel_loss_1}
  \end{minipage}
  \begin{minipage}[b]{0.45\linewidth}
    \centering
    \includegraphics[keepaspectratio, scale=0.4]{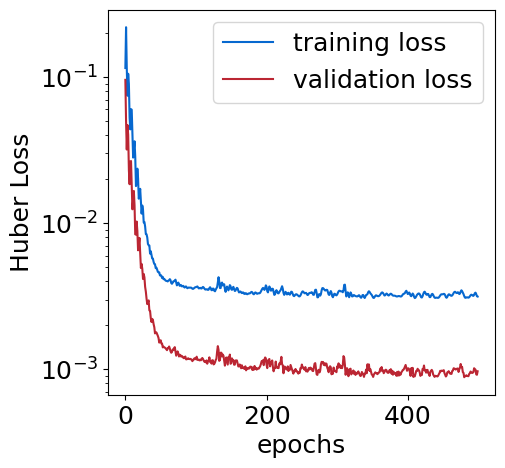}
    \subcaption{loss function of \(\mathcal{M}_{\mathrm{t},~\mathrm{IDR}}\) trained by the dataset that include \(20\) samples}\label{fig:cs1_disp_loss_1}
  \end{minipage}
  \caption{Loss functions of \(\mathcal{M}_\mathrm{t}\) in training}
  \label{fig:cs1loss}
  \centering
    \includegraphics[width=0.9\columnwidth]{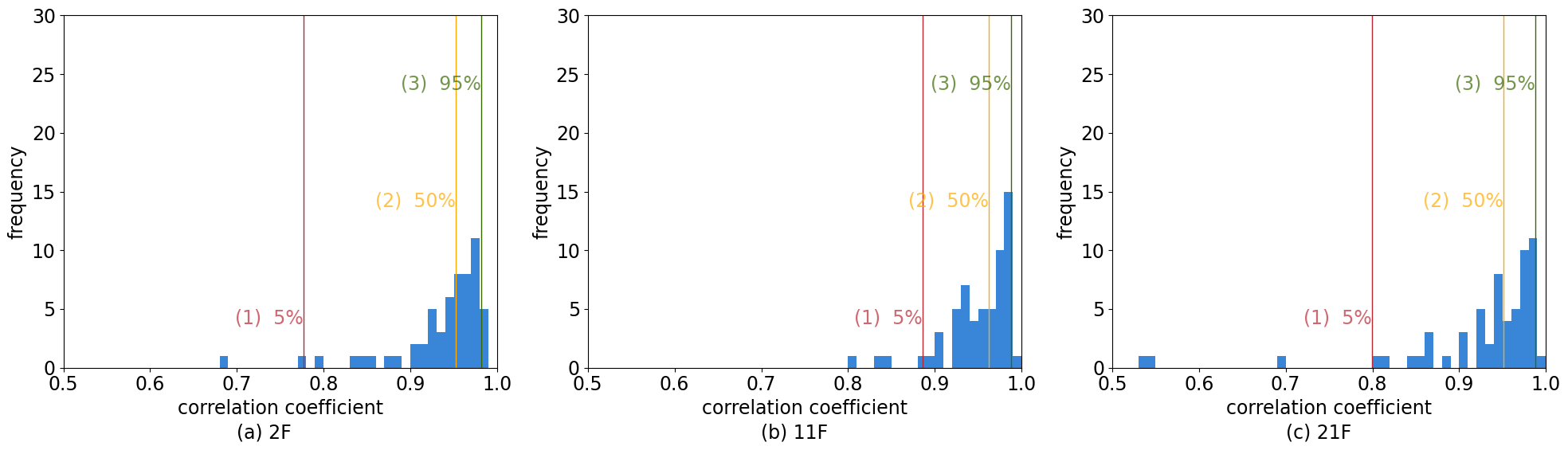}
    \caption{Distribution of correlation coefficients between response analyses results and prediction by \(\mathcal{M}_{\mathrm{t},~\mathrm{accel}}\)}
    \label{fig:cs1_accel_MSE}
  \centering
  \vspace{5mm}
  \centering
    \includegraphics[width=0.9\columnwidth]{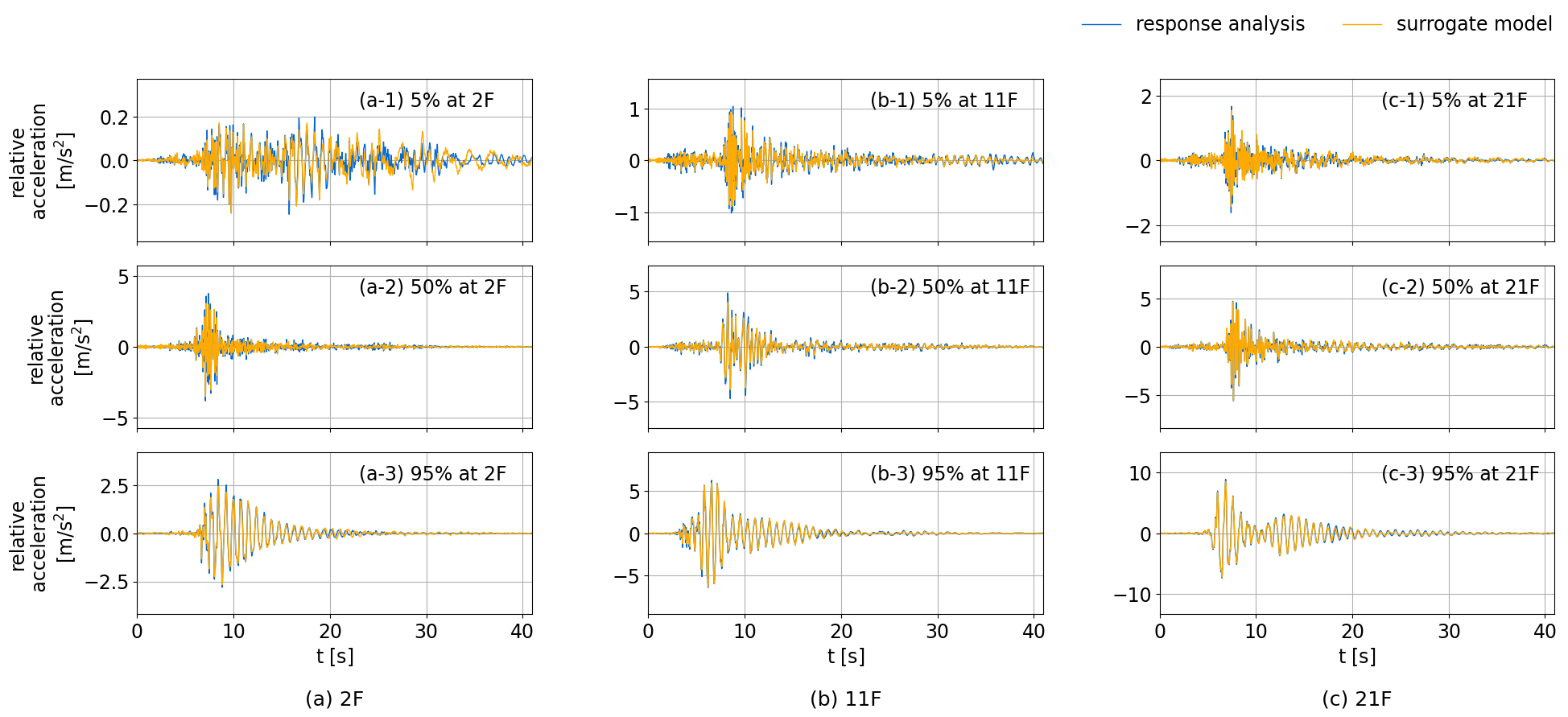}
    \caption{Comparison of response analysis results and prediction by \(\mathcal{M}_{\mathrm{t},~\mathrm{accel}}\)}
    \label{fig:cs1_accel_res}
\end{figure}

\begin{align}
    r_{i, j} = \frac{\mathrm{Cov}\left(\mathbf{y}_{i, j}, \hat{\mathbf{y}}_{i, j}\right)}{\sqrt{\mathrm{Var}\left(\mathbf{y}_{i, j}\right) \mathrm{Var}\left(\hat{\mathbf{y}}_{i, j}\right)}}
\end{align}
where \(\mathrm{Cov}\) and \(\mathrm{Var}\) denote covariance and variance, respectively; \(i\) represents the sample index; and \(j\) represents the type of response (relative acceleration, relative velocity, relative displacement, or lateral force of the shear spring).
The distribution of correlation coefficients for each type of response to \(60\) ground motions selected for the validation dataset is shown in Figure~\ref{fig:cs1_MS_valid_dist}.
The red, orange, and green vertical lines in Figure~\ref{fig:cs1_MS_valid_dist} represent the bottom 5th percentile (1), 50th percentile (2), and 95th percentile (3), respectively.
The samples at these percentiles are shown in Figure~\ref{fig:cs1_MS_valid_wave}.
The results demonstrate that the trained \(\mathcal{M}_\mathrm{s}\) can accurately predict the responses of \(\mathcal{R}_\mathrm{s}\).

\begin{figure}[t]
  \centering
    \includegraphics[width=0.9\columnwidth]{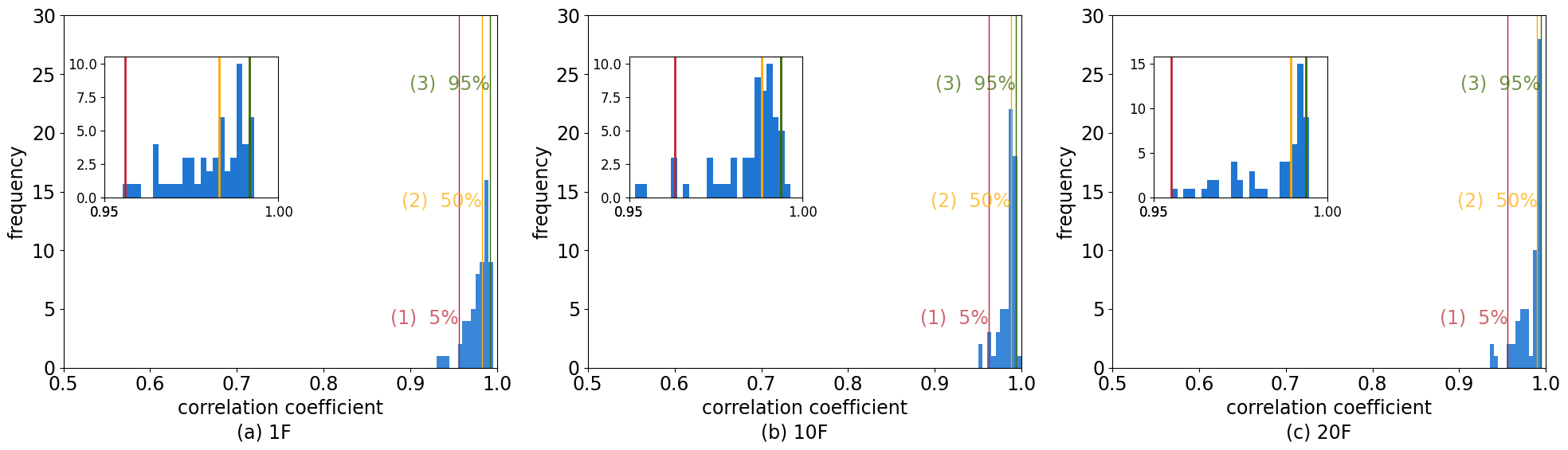}
    \caption{Distribution of correlation coefficients between response analyses results and prediction by \(\mathcal{M}_{\mathrm{t},~\mathrm{IDR}}\)}
    \label{fig:cs1_SDR_MSE}
  \centering
  \vspace{5mm}
  \includegraphics[width=0.9\columnwidth]{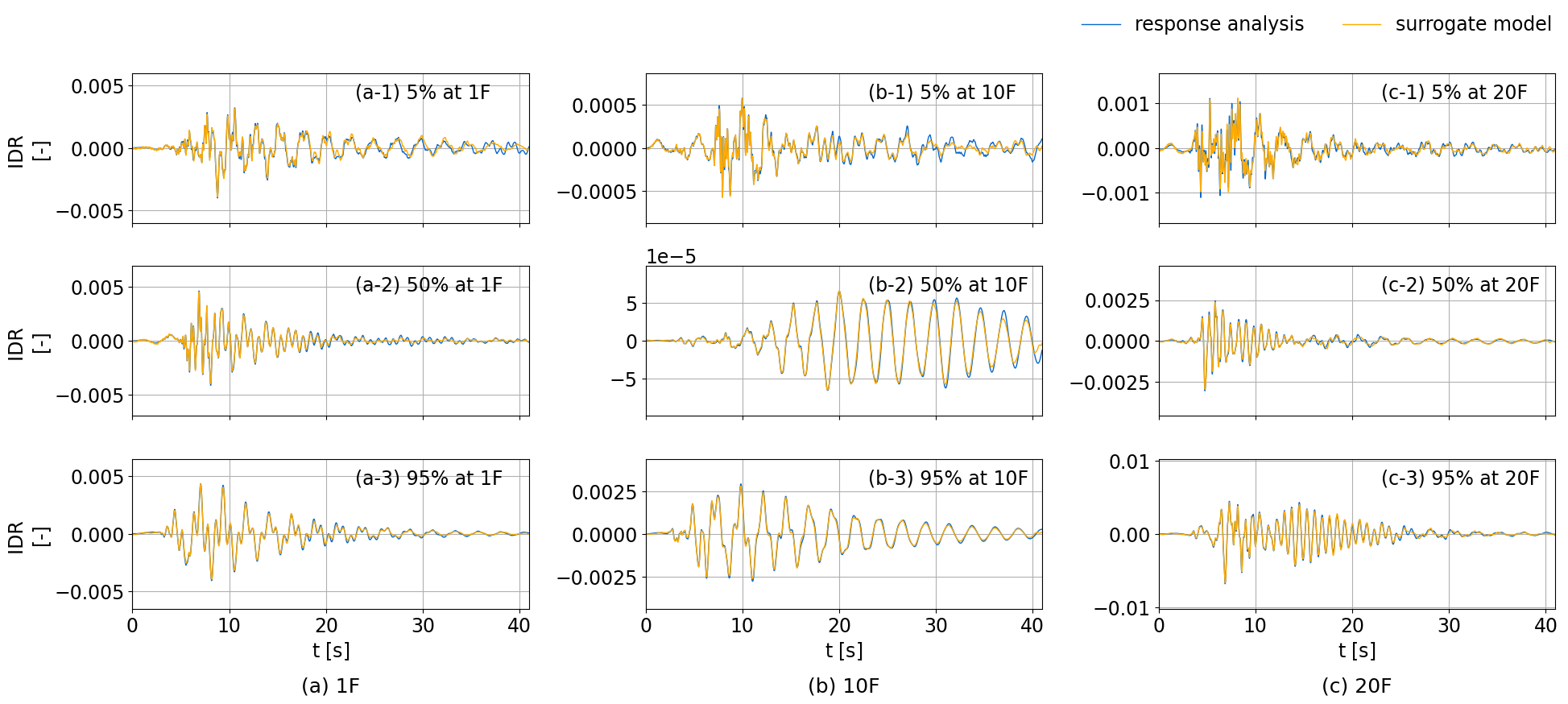}
    \caption{Comparison of response analysis results and prediction by \(\mathcal{M}_{\mathrm{t},~\mathrm{IDR}}\)}
    \label{fig:cs1_SDR_res}
\end{figure}

Training of the two target models \(\mathcal{M}_{\mathrm{t}, \mathrm{accel}}\) and \(\mathcal{M}_{\mathrm{t}, \mathrm{IDR}}\) were then conducted using the Adam optimizer with a learning rate of \(1.0 \times 10^{-3}\).
The loss functions over the training epochs are shown in Figure~\ref{fig:cs1loss}, for the case of training data size \(N = 20\).
To validate the prediction performance of \(\mathcal{M}_\mathrm{t}\), the correlation coefficient between ground-truth and predicted responses was computed as
\begin{align}
    r_{i, j, k} = \frac{\mathrm{Cov}\left(\mathbf{y}_{i, j, k}, \hat{\mathbf{y}}_{i, j, k}\right)}{\sqrt{\mathrm{Var}\left(\mathbf{y}_{i, j, k}\right), \mathrm{Var}\left(\hat{\mathbf{y}}_{i, j, k}\right)}}
\end{align}
where \(j\) represents the type of response (relative acceleration or IDR), and \(k\) represents the floor index.
The distributions of \(r_{i,~\mathrm{accel}, ~k}\) and \(r_{i,~\mathrm{IDR},~k}\) are shown in Figures~\ref{fig:cs1_accel_MSE} and~\ref{fig:cs1_SDR_MSE}, respectively.
The red, orange, and green vertical lines in Figures~\ref{fig:cs1_accel_MSE} and~\ref{fig:cs1_SDR_MSE} correspond to the top \(\mbox{(1)}~5 \%, \mbox{(2)}~50\%, ~\mbox{and (3)}~95\%\), respectively.
Figures~\ref{fig:cs1_accel_res} and~\ref{fig:cs1_SDR_res} compare the ground-truth and predicted responses at these percentiles, demonstrating that both \(\mathcal{M}_{\mathrm{t,accel}}\) and \(\mathcal{M}_\mathrm{t, IDR}\) can predict time-history responses accurately.

\begin{figure}[b!]
\centering
    \begin{minipage}[b]{0.45\linewidth}
    \centering
    \includegraphics[keepaspectratio, scale=0.3]{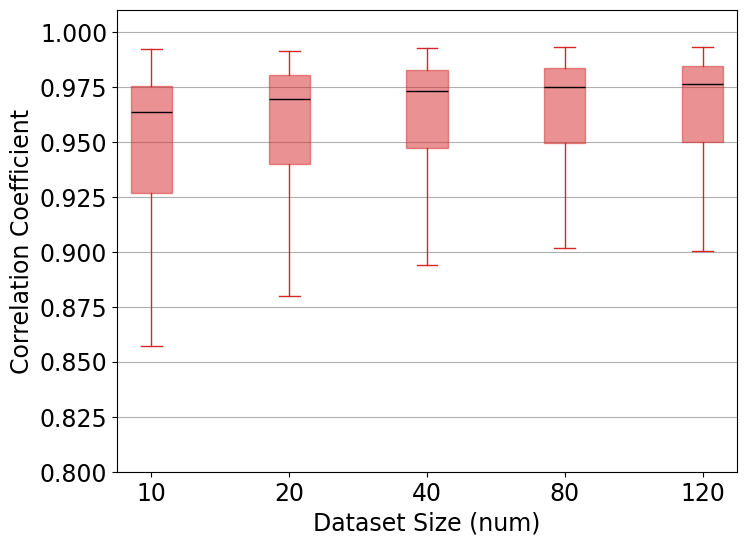}
    \subcaption{Distribution of the average of correlation coefficients of \(\mathcal{M}_{\mathrm{t}, ~\mathrm{accel}}\)}\label{fig:cs1_loss_dist_accel}
  \end{minipage}
  \begin{minipage}[b]{0.45\linewidth}
    \centering
    \includegraphics[keepaspectratio, scale=0.3]{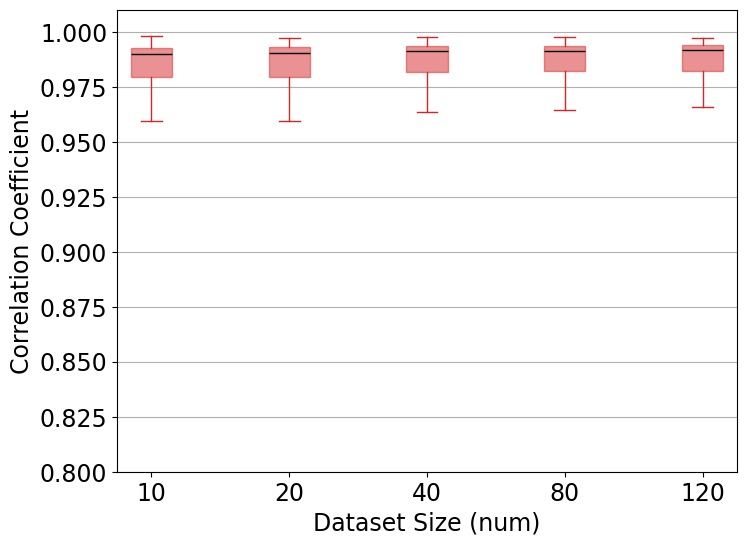}
    \subcaption{Distribution of the average of correlation coefficients of  \(\mathcal{M}_{\mathrm{t}, ~\mathrm{IDR}}\)}\label{fig:cs1_loss_dist_SDR}
  \end{minipage}
  \caption{Distributions of average of correlation coefficients between response analysis result and prediction by \(\mathcal{M}_\mathrm{t}\) to ground motions selected for validation datasets}
  \label{fig:cs1loss_dist}
\centering
\begin{minipage}[b]{0.45\linewidth}
    \centering
    \includegraphics[keepaspectratio, scale=0.4]{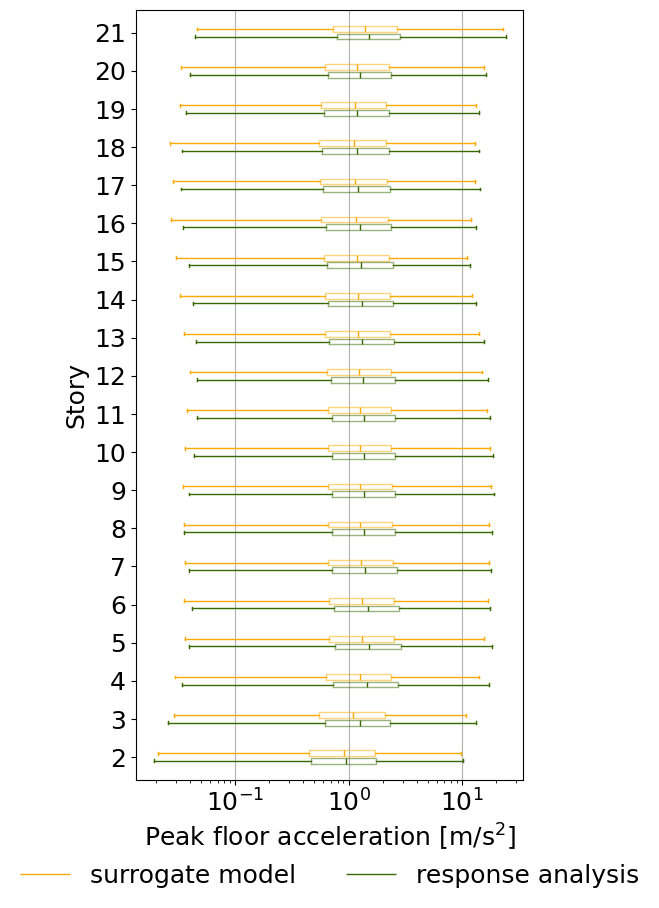}
    \subcaption{The distribution of peak floor acceleration at each floor}\label{fig:cs1_accel_dist}
  \end{minipage}
  \begin{minipage}[b]{0.45\linewidth}
    \centering
    \includegraphics[keepaspectratio, scale=0.4]{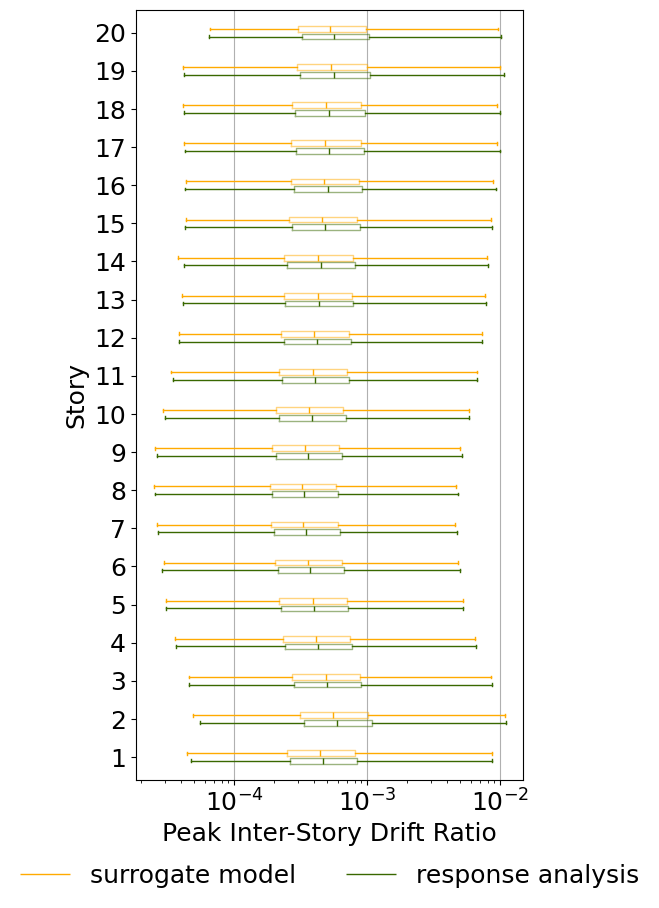}
    \subcaption{The distribution of inter-story drift ratio at each floor}\label{fig:cs1_SDR_dist}
  \end{minipage}
  \caption{Comparison of responses to the \(10,000\) ground motions selected randomly from the hazard calculated by response analysis and predicted by \(\mathcal{M}_\mathrm{t}\)}
  \label{fig:cs1_resp_dist}
\end{figure}

We then examined the effects of the training dataset size \(N\) on the prediction performance of \(\mathcal{M}_{\mathrm{t}}\).
Nine independent runs of FPS were conducted for each value of \(N \in \{10, 20, 40, 80, 120\}\) by changing the random seed, which resulted in a total of $9 \times 5 = 45$ training datasets for each of \(\mathcal{M}_{\mathrm{t},\mathrm{accel}}\) and \(\mathcal{M}_{\mathrm{t}, \mathrm{IDR}}\).
%
For each case, the predictive accuracy was quantified using the 60 validation ground motions.
The correlation coefficient (\(r_{i, j, k}\)) between the predicted and true time-history responses was calculated for all 20 floors~(\(k\)) and all \(60\) validation samples (\(i\)).
To aggregate the floor-wise results, an average correlation coefficient \(\bar{r}_{i, j}\) was calculated for each validation ground motion (\(i\)) across all 20 floors~(\(k\)) for a specific training seed:
\begin{align}
    \bar{r}_{i,j} = \frac{1}{20}\sum_{k=1}^{20}r_{i, j, k}
\end{align}
For each dataset size (\(N\)) and each type of responses (floor acceleration and IDR), the box plots shown in Figure~\ref{fig:cs1loss_dist} were constructed from a total of \(9\times60=540\) individual average correlation values (\(\bar{r}_{i, j}\)).
The whiskers in the box plots extend to the most extreme data points not considered outliers, typically within \(1.5\) times the inter-quartile range (IQR).
This provides a robust visualization of the variability in model performance across both different training dataset sizes and multiple training initializations (random seeds).

\begin{figure}[p]
\centering
    \includegraphics[width=0.9\columnwidth]{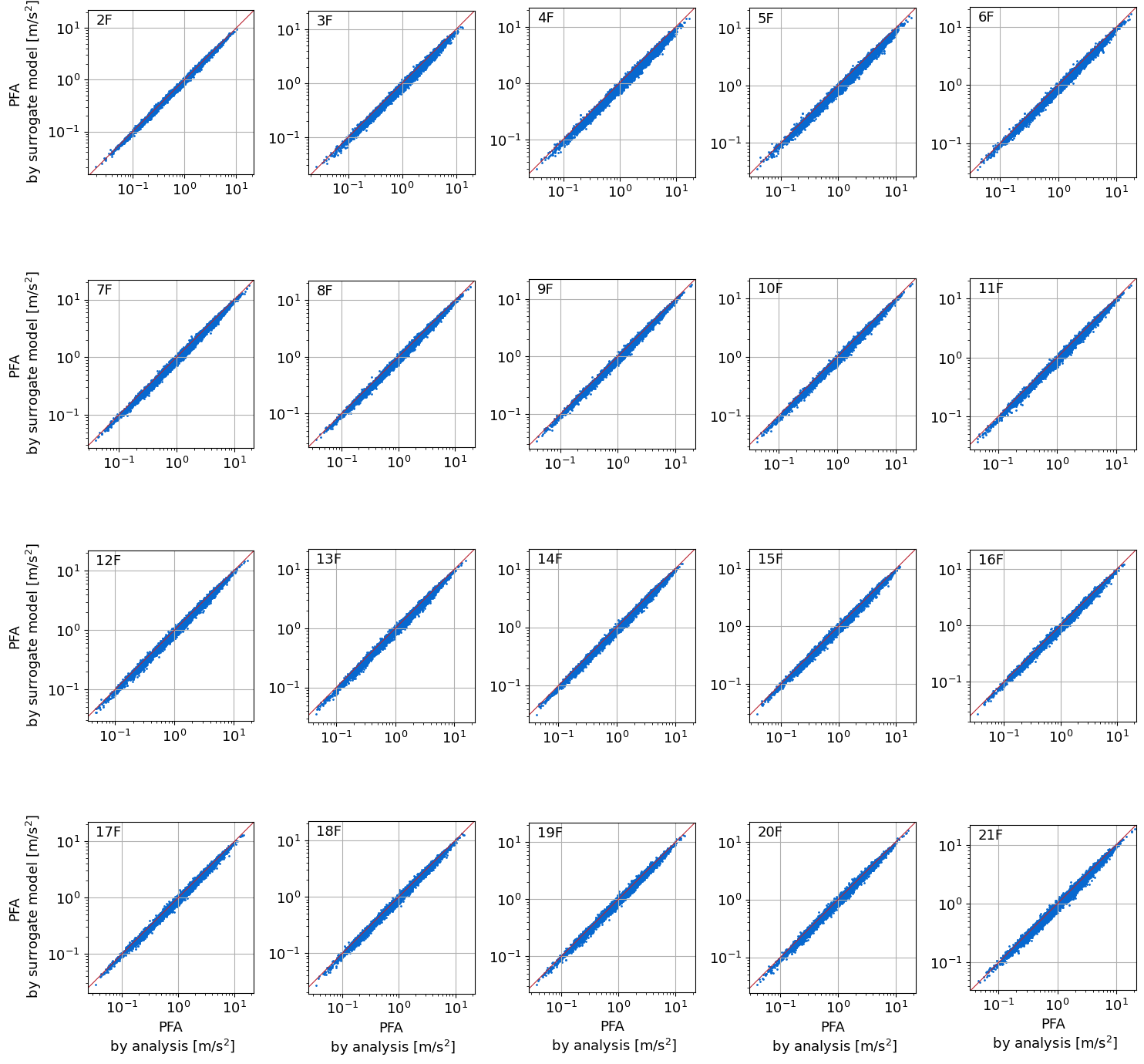}
    \caption{The comparison of peak floor acceleration to the \(10000\) ground motions selected randomly from the hazard calculated by response analysis and predicted by \(\mathcal{M}_\mathrm{t}\)}
    \label{fig:cs1_accel_max_log}
\end{figure}
\begin{figure}[p]
\centering
    \includegraphics[width=0.9\columnwidth]{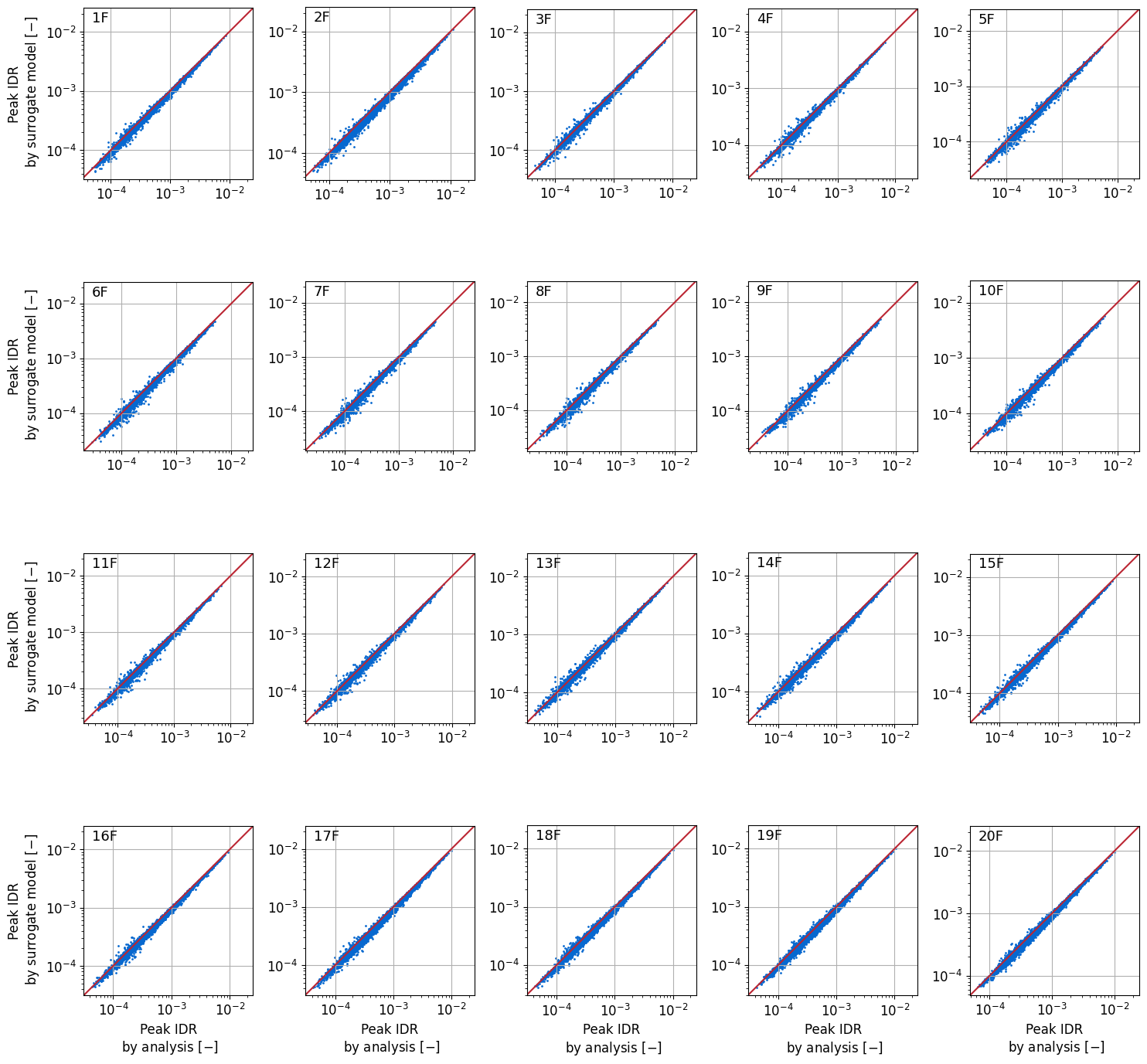}
    \caption{The comparison of peak inter-story drift ratio (IDR) to the \(10000\) ground motions selected randomly from the hazard calculated by response analysis and predicted by \(\mathcal{M}_\mathrm{t}\)}
    \label{fig:cs1_SDR_max_log}
\end{figure}
\begin{figure}
\centering
    \begin{minipage}[b]{0.45\linewidth}
    \centering
    \includegraphics[keepaspectratio, scale=0.4]{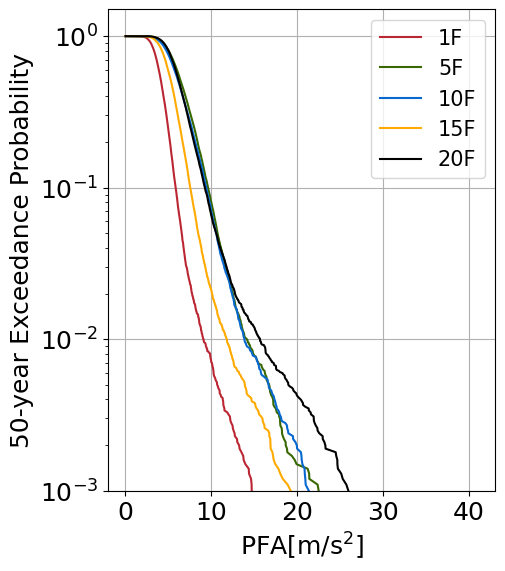}
    \subcaption{The 50-year exceedance probability of peak floor acceleration estimated with \(\mathcal{M}_{\mathrm{t}, ~\mathrm{accel}}\)}\label{fig:EP_PFA.png}
  \end{minipage}
  \begin{minipage}[b]{0.45\linewidth}
    \centering
    \includegraphics[keepaspectratio, scale=0.4]{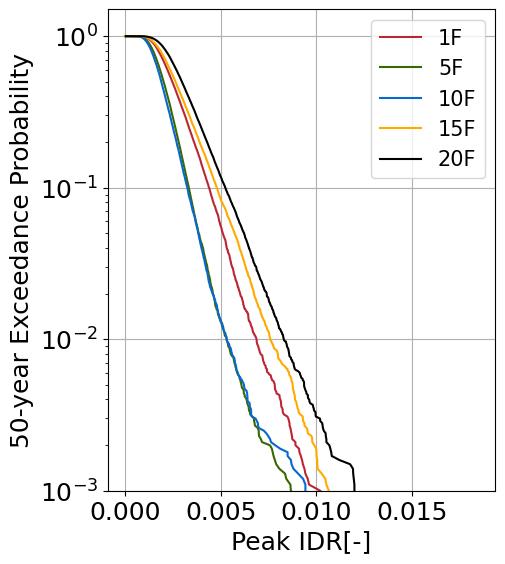}
    \subcaption{The 50-year exceedance probability of peak IDR estimated with \(\mathcal{M}_{\mathrm{t}, ~\mathrm{IDR}}\)}\label{fig:EP_IDR.png}
  \end{minipage}
  \caption{The exceedance probability of PFA and peak IDR at 1st, 5th, 10th, 15th, 20th floor}
  \label{fig:EP}
\end{figure}

\subsection{Assessment of Exceedance Probability of Hazard-Consistent Engineering Demand Parameters}\label{sec:comp_surro_res}
To demonstrate that \(\mathcal{M}_\mathrm{t}\) can accurately predict EDPs of \(\mathcal{R}_\mathrm{t}\) for the ground motions within the considered hazard, \(10,000\) ground motions were selected randomly from the entire \(250,476\) ground motions in the hazard.
NLTHAs were conducted for each selected ground motion, and prediction of responses by \(\mathcal{M}_{\mathrm{t}, \mathrm{accel}}\) and \(\mathcal{M}_{\mathrm{t}, \mathrm{IDR}}\) to the selected ground motions were also conducted.
The distributions of the ground-truth and predicted EDPs (PFA and peak IDR) on all floors are shown using box plots in Figures~\ref{fig:cs1_accel_dist} and \ref{fig:cs1_SDR_dist}.
These results confirm that \(\mathcal{M}_\mathrm{t}\), trained with the proposed framework, can accurately predict statistical distribution of responses to ground motions contained in the hazard with the predicted distribution exhibiting a slight underestimation compared to the ground truth.
Additionally, to visualize the accuracy of each prediction, PFA and peak IDR of each ground motion are plotted in Figures~\ref{fig:cs1_accel_max_log} and~\ref{fig:cs1_SDR_max_log}.
These results demonstrate that \(\mathcal{M}_\mathrm{t}\) accurately predicts not only distribution of the responses but also the peak value of responses to each ground motions.

To demonstrate the practical capability of the constructed surrogate models \(\mathcal{M}_\mathrm{t}\), we assessed the 50-year exceedance probability of PFA and peak IDR, which were predicted using \(\mathcal{M}_\mathrm{t}\) for the entire \(250,476\) ground motions in the generated suite were predicted using \(\mathcal{M}_\mathrm{t}\).
The generated suite was consistent with seismic hazard at the site during an evaluation period of 50 years.
The resulting exceedance probabilities are plotted in Figure~\ref{fig:EP}.
By incorporating the proposed framework, probabilistic evaluation of structural responses based on high-fidelity models—which was previously computationally prohibitive—becomes feasible.

\section{Conclusions and Future Challenges}
In this paper, we propose a transfer learning-based framework that enables the efficient construction of surrogate models for high-fidelity response analysis models with a limited number of numerical simulations as training data.
The proposed framework employs low-fidelity response analysis models as source-domain models and applies transfer learning, which allows the development of the target surrogate models without large training datasets.
Utilizing the proposed framework and newly developed MNN architecture, in which causality is incorporated, we successfully constructed a surrogate model for a 20-story SMF.
An SDOF model was used as the low-fidelity response analysis model, and the surrogate model of the SMF model was constructed with as few as 20 training samples.
The constructed surrogate models can accurately predict time-history responses of the SMF and representative values (e.g., PFA and IDR) for hazard-consistent ground motions generated by the machine learning-based ground motion generative model (GMGM)\cite{matsumoto2025}.
In conclusion, the newly developed framework provides an efficient and robust method for constructing high-fidelity structural surrogate models which are practical for performance evaluation of buildings and civil engineering structures.

Although this study enables the efficient construction of surrogate models for higher-fidelity structural time-history responses, current damage assessment of buildings and components still largely relies on peak response metrics, such as peak floor acceleration (PFA) and peak inter-story drift ratio (IDR).
Further studies are therefore required to develop fragility databases that explicitly incorporate time-history response information. Such developments would allow the full capabilities of the proposed surrogate modeling framework to be leveraged and would further enhance the PBEE framework.
In addition, while the present study focuses on a single representative steel moment frame, the proposed framework is not restricted to this configuration and can be applied to other structural systems and building scales, which will be reported in future studies.

\section*{Declaration of generative AI and AI-assisted technologies in the manuscript preparation process}


During the preparation of this work the authors used ChatGPT and Gemini in order to improve the readability and language of the manuscript.
After using this tool, the authors reviewed and edited the content as needed and take full responsibility for the content of the published article.

\bibliographystyle{unsrt}
\bibliography{bib}

@article{Soleimani2022,
author = {Soleimani, Farahnaz and Liu, Xi},
title = {Artificial neural network application in predicting probabilistic seismic demands of bridge components},
journal = {Earthquake Engineering \& Structural Dynamics},
volume = {51},
number = {3},
pages = {612-629},
doi = {https://doi.org/10.1002/eqe.3582},
url = {https://onlinelibrary.wiley.com/doi/abs/10.1002/eqe.3582},
eprint = {https://onlinelibrary.wiley.com/doi/pdf/10.1002/eqe.3582},
year = {2022}
}

@article{Chara2011,
title = {Developing fragility curves based on neural network IDA predictions},
journal = {Engineering Structures},
volume = {33},
number = {12},
pages = {3409-3421},
year = {2011},
issn = {0141-0296},
doi = {https://doi.org/10.1016/j.engstruct.2011.07.005},
url = {https://www.sciencedirect.com/science/article/pii/S0141029611002756},
author = {Chara Ch. Mitropoulou and Manolis Papadrakakis}
}

@article{ZHANG201955,
title = {Deep long short-term memory networks for nonlinear structural seismic response prediction},
journal = {Computers \& Structures},
volume = {220},
pages = {55-68},
year = {2019},
issn = {0045-7949},
doi = {https://doi.org/10.1016/j.compstruc.2019.05.006},
url = {https://www.sciencedirect.com/science/article/pii/S0045794919302263},
author = {Ruiyang Zhang and Zhao Chen and Su Chen and Jingwei Zheng and Oral Büyüköztürk and Hao Sun}
}

@article{Ni2022,
AUTHOR = {Ni, Peng and Sun, Limin and Yang, Jipeng and Li, Yixian},
TITLE = {Multi-End Physics-Informed Deep Learning for Seismic Response Estimation},
JOURNAL = {Sensors},
VOLUME = {22},
YEAR = {2022},
NUMBER = {10},
ARTICLE-NUMBER = {3697},
URL = {https://www.mdpi.com/1424-8220/22/10/3697},
PubMedID = {35632106},
ISSN = {1424-8220},
DOI = {10.3390/s22103697}
}

@article{Zhong2023,
author = {Zhong, Kuanshi and Navarro, Javier G. and Govindjee, Sanjay and Deierlein, Gregory G.},
title = {Surrogate modeling of structural seismic response using probabilistic learning on manifolds},
journal = {Earthquake Engineering \& Structural Dynamics},
volume = {52},
number = {8},
pages = {2407-2428},
doi = {https://doi.org/10.1002/eqe.3839},
url = {https://onlinelibrary.wiley.com/doi/abs/10.1002/eqe.3839},
eprint = {https://onlinelibrary.wiley.com/doi/pdf/10.1002/eqe.3839},
year = {2023}
}

@article{Bojórquez2013,
author = {Bojórquez, Edén and Ruiz-García, Jorge},
title = {Residual drift demands in moment-resisting steel frames subjected to narrow-band earthquake ground motions},
journal = {Earthquake Engineering \& Structural Dynamics},
volume = {42},
number = {11},
pages = {1583-1598},
doi = {https://doi.org/10.1002/eqe.2288},
url = {https://onlinelibrary.wiley.com/doi/abs/10.1002/eqe.2288},
eprint = {https://onlinelibrary.wiley.com/doi/pdf/10.1002/eqe.2288},
year = {2013}
}

@article{NING2023,
title = {LSTM, WaveNet, and 2D CNN for nonlinear time history prediction of seismic responses},
journal = {Engineering Structures},
volume = {286},
pages = {116083},
year = {2023},
issn = {0141-0296},
doi = {https://doi.org/10.1016/j.engstruct.2023.116083},
url = {https://www.sciencedirect.com/science/article/pii/S0141029623004972},
author = {Chunxiao Ning and Yazhou Xie and Lijun Sun}
}

@article{SAIDA2025120953,
title = {ExSRNet: Explainable deep learning model for seismic response prediction with frequency attention mechanism},
journal = {Engineering Structures},
volume = {343},
pages = {120953},
year = {2025},
issn = {0141-0296},
doi = {https://doi.org/10.1016/j.engstruct.2025.120953},
url = {https://www.sciencedirect.com/science/article/pii/S0141029625013446},
author = {Taisei Saida and Mayuko Nishio}
}

@article{ZHU20186,
title = {OpenSeesPy: Python library for the OpenSees finite element framework},
journal = {SoftwareX},
volume = {7},
pages = {6-11},
year = {2018},
issn = {2352-7110},
doi = {https://doi.org/10.1016/j.softx.2017.10.009},
url = {https://www.sciencedirect.com/science/article/pii/S2352711017300584},
author = {Minjie Zhu and Frank McKenna and Michael H. Scott}
}

@article{ZHANG2020,
title = {Physics-guided convolutional neural network (PhyCNN) for data-driven seismic response modeling},
journal = {Engineering Structures},
volume = {215},
pages = {110704},
year = {2020},
issn = {0141-0296},
doi = {https://doi.org/10.1016/j.engstruct.2020.110704},
url = {https://www.sciencedirect.com/science/article/pii/S0141029619345080},
author = {Ruiyang Zhang and Yang Liu and Hao Sun}
}

@book{NIST2010,
  author = {Charles Kircher and Gregory Deierlein and John Hooper and Helmut Krawinkler and Steve Mahin and Benson Shing and John Wallace},
  publisher = {Grant Contract Reports (NISTGCR), National Institute of Standards and Technology, Gaithersburg, MD},
  url = {https://tsapps.nist.gov/publication/get_pdf.cfm?pub_id=915492},
  language = {en},
  title = {Evaluation of the FEMA P-695 Methodology for Quantification of Building Seismic Performance Factors},
  year = {2010},
  month = {2010-11-15},
}

@article{Ibarra2005,
author = {Ibarra, Luis and Medina, Ricardo and Krawinkler, Helmut},
year = {2005},
month = {10},
pages = {1489 - 1511},
title = {Hysteretic Models that Incorporate Strength and Stiffness Deterioration},
volume = {34},
journal = {Earthquake Engineering \& Structural Dynamics},
doi = {10.1002/eqe.495}
}

@article{Lignos2011,
author = {Dimitrios G. Lignos  and Helmut Krawinkler },
title = {Deterioration Modeling of Steel Components in Support of Collapse Prediction of Steel Moment Frames under Earthquake Loading},
journal = {Journal of Structural Engineering},
volume = {137},
number = {11},
pages = {1291-1302},
year = {2011},
doi = {10.1061/(ASCE)ST.1943-541X.0000376},

URL = {https://ascelibrary.org/doi/abs/10.1061/ST.1943-541X.0000376},
eprint = {https://ascelibrary.org/doi/pdf/10.1061/ST.1943-541X.0000376}
}

@misc{peer2010,
title = {Modeling and Acceptance Criteria for Seismic Design and Analysis of Tall Buildings},
year = {2010},
note = {PEER Report 2010-111},
author = {{Pacific Earthquake Engineering Research Center}},
note = {University of California, Berkeley, CA}
}

@article{FS1997,
author = {Eldar, Yuval and Lindenbaum, Michael and Porat, Moshe and Zeevi, Yehoshua},
year = {1997},
month = {02},
pages = {1305-15},
title = {The Farthest Point Strategy for Progressive Image Sampling},
volume = {6},
journal = {IEEE Transactions on Image Processing},
doi = {10.1109/83.623193}
}

@article{Cornell1968,
    author = {Cornell, C. Allin},
    title = {Engineering seismic risk analysis},
    journal = {Bulletin of the Seismological Society of America},
    volume = {58},
    number = {5},
    pages = {1583-1606},
    year = {1968},
    month = {10},
    issn = {0037-1106},
    doi = {10.1785/BSSA0580051583},
    url = {https://doi.org/10.1785/BSSA0580051583},
}

@article{Gunay18082013,
author = {Selim Günay and Khalid M. Mosalam},
title = {PEER Performance-Based Earthquake Engineering Methodology, Revisited},
journal = {Journal of Earthquake Engineering},
volume = {17},
number = {6},
pages = {829--858},
year = {2013},
publisher = {Taylor \& Francis},
doi = {10.1080/13632469.2013.787377},
URL = {https://doi.org/10.1080/13632469.2013.787377},
eprint = {https://doi.org/10.1080/13632469.2013.787377}
}

@misc{FEMAp58,
author = {FEMA},
year = {2018},
month = {09},
title = {Seismic Performance Assessment of Buildings, Volume 1 – Methodology Second Edition, FEMA P-58-1},
url={https://femap58.atcouncil.org/reports}
}

@inproceedings{Moehle2004AFM,
  title={A framework methodology for performance-based earthquake engineering},
  author={Jack P. Moehle and Gregory G. Deierlein},
  year={2004},
  url={https://api.semanticscholar.org/CorpusID:107891620}
}

@article{BAI201196,
title = {Story-specific demand models and seismic fragility estimates for multi-story buildings},
journal = {Structural Safety},
volume = {33},
number = {1},
pages = {96-107},
year = {2011},
issn = {0167-4730},
doi = {https://doi.org/10.1016/j.strusafe.2010.09.002},
url = {https://www.sciencedirect.com/science/article/pii/S0167473010000809},
author = {Jong-Wha Bai and Paolo Gardoni and Mary Beth D. Hueste},
}

@article{Jack2011,
author = {Jack W. Baker },
title = {Conditional Mean Spectrum: Tool for Ground-Motion Selection},
journal = {Journal of Structural Engineering},
volume = {137},
number = {3},
pages = {322-331},
year = {2011},
doi = {10.1061/(ASCE)ST.1943-541X.0000215},

URL = {https://ascelibrary.org/doi/abs/10.1061/ST.1943-541X.0000215},
eprint = {https://ascelibrary.org/doi/pdf/10.1061/ST.1943-541X.0000215}
}

@misc{matsumoto2025,
      title={Waveform-Based Probabilistic Seismic Hazard Analysis Using Ground-Motion Generative Models}, 
      author={Yuma Matsumoto and Taro Yaoyama and Sangwon Lee and Asako Iwaki and Tatsuya Itoi},
      year={2025},
      eprint={2511.22106},
      archivePrefix={arXiv},
      primaryClass={physics.geo-ph},
      url={https://arxiv.org/abs/2511.22106}, 
}

@article{Matsumoto2024,
    author = {Matsumoto, Yuma and Yaoyama, Taro and Lee, Sangwon and Hida, Takenori and Itoi, Tatsuya},
    title = {Generative Adversarial Networks‐Based Ground‐Motion Model for Crustal Earthquakes in Japan Considering Detailed Site Conditions},
    journal = {Bulletin of the Seismological Society of America},
    volume = {114},
    number = {6},
    pages = {2886-2911},
    year = {2024},
    month = {09},
    issn = {0037-1106},
    doi = {10.1785/0120240070},
    url = {https://doi.org/10.1785/0120240070},
    eprint = {https://pubs.geoscienceworld.org/ssa/bssa/article-pdf/114/6/2886/7055798/bssa-2024070.1.pdf},
}

@article{Matsumoto2023,
author = {Matsumoto, Yuma and Yaoyama, Taro and Lee, Sangwon and Hida, Takenori and Itoi, Tatsuya},
title = {Fundamental study on probabilistic generative modeling of earthquake ground motion time histories using generative adversarial networks},
journal = {Japan Architectural Review},
volume = {6},
number = {1},
pages = {e12392},
doi = {https://doi.org/10.1002/2475-8876.12392},
url = {https://onlinelibrary.wiley.com/doi/abs/10.1002/2475-8876.12392},
eprint = {https://onlinelibrary.wiley.com/doi/pdf/10.1002/2475-8876.12392},
note = {e12392 JAR-2023-0055.R2},
year = {2023}
}

@article{Esfahani2022,
    author = {Esfahani, Reza D. D. and Cotton, Fabrice and Ohrnberger, Matthias and Scherbaum, Frank},
    title = {TFCGAN: Nonstationary Ground‐Motion Simulation in the Time–Frequency Domain Using Conditional Generative Adversarial Network (CGAN) and Phase Retrieval Methods},
    journal = {Bulletin of the Seismological Society of America},
    volume = {113},
    number = {1},
    pages = {453-467},
    year = {2022},
    month = {12},
    issn = {0037-1106},
    doi = {10.1785/0120220068},
    url = {https://doi.org/10.1785/0120220068},
    eprint = {https://pubs.geoscienceworld.org/ssa/bssa/article-pdf/113/1/453/5770813/bssa-2022068.1.pdf},
}

@article{Florez2022,
    author = {Florez, Manuel A. and Caporale, Michaelangelo and Buabthong, Pakpoom and Ross, Zachary E. and Asimaki, Domniki and Meier, Men‐Andrin},
    title = {Data‐Driven Synthesis of Broadband Earthquake Ground Motions Using Artificial Intelligence},
    journal = {Bulletin of the Seismological Society of America},
    volume = {112},
    number = {4},
    pages = {1979-1996},
    year = {2022},
    month = {04},
    issn = {0037-1106},
    doi = {10.1785/0120210264},
    url = {https://doi.org/10.1785/0120210264},
    eprint = {https://pubs.geoscienceworld.org/ssa/bssa/article-pdf/112/4/1979/5659704/bssa-2021264.1.pdf},
}

@article{Yamaguchi2024,
    author = {Yamaguchi, Junki and Tomozawa, Yusuke and Saka, Toshihide},
    title = {Site‐Specific Ground‐Motion Waveform Generation Using a Conditional Generative Adversarial Network and Generalized Inversion Technique},
    journal = {Bulletin of the Seismological Society of America},
    volume = {114},
    number = {4},
    pages = {2118-2137},
    year = {2024},
    month = {04},
    issn = {0037-1106},
    doi = {10.1785/0120230209},
    url = {https://doi.org/10.1785/0120230209},
    eprint = {https://pubs.geoscienceworld.org/ssa/bssa/article-pdf/114/4/2118/6586548/bssa-2023209.1.pdf},
}

@article{Shi2024,
    author = {Shi, Yaozhong and Lavrentiadis, Grigorios and Asimaki, Domniki and Ross, Zachary E. and Azizzadenesheli, Kamyar},
    title = {Broadband Ground‐Motion Synthesis via Generative Adversarial Neural Operators: Development and Validation},
    journal = {Bulletin of the Seismological Society of America},
    volume = {114},
    number = {4},
    pages = {2151-2171},
    year = {2024},
    month = {03},
    issn = {0037-1106},
    doi = {10.1785/0120230207},
    url = {https://doi.org/10.1785/0120230207},
    eprint = {https://pubs.geoscienceworld.org/ssa/bssa/article-pdf/114/4/2151/6586500/bssa-2023207.1.pdf},
}

@misc{dyneq,
author="Yoshida, N",
title="DYNEQ: a computer program for dynamic analysis of level ground based on equivalent linear method",
journal="Reports of Engineering Research Institute",
year="1996",
pages="61",
URL="https://cir.nii.ac.jp/crid/1370013168779924110"
}

@inproceedings{optuna_2019,
    title={Optuna: A Next-generation Hyperparameter Optimization Framework},
    author={Akiba, Takuya and Sano, Shotaro and Yanase, Toshihiko and Ohta, Takeru and Koyama, Masanori},
    booktitle={Proceedings of the 25th {ACM} {SIGKDD} International Conference on Knowledge Discovery and Data Mining},
    year={2019}
}

@book{AIJ2006,
  author = {{Architectural Institute of Japan}},
  year = {2006},
  title = {Seismic Response Analysis and Design of Buildings Considering Dynamic Soil-Structure Interaction},
  publisher = {Maruzen},
  note={(in Japanese)}
}

@ARTICLE{zuhangTL,
  author={Zhuang, Fuzhen and Qi, Zhiyuan and Duan, Keyu and Xi, Dongbo and Zhu, Yongchun and Zhu, Hengshu and Xiong, Hui and He, Qing},
  journal={Proceedings of the IEEE}, 
  title={A Comprehensive Survey on Transfer Learning}, 
  year={2021},
  volume={109},
  number={1},
  pages={43-76},
  doi={10.1109/JPROC.2020.3004555}}

@article{Wang2011,
  author={Wang, Hua and Feiping Nie and Huang, Heng and Ding, Chris},
  booktitle={2011 International Conference on Computer Vision}, 
  title={Dyadic transfer learning for cross-domain image classification}, 
  year={2011},
  pages={551-556},
  doi={10.1109/ICCV.2011.6126287}}


\end{document}